\documentclass[11pt,a4paper]{article} 
\usepackage[utf8]{inputenc}
\usepackage{latexsym}
\usepackage{amssymb}
\usepackage{amsmath}
\usepackage{amsthm}
\usepackage{stmaryrd}
\usepackage{graphicx}
\usepackage{bm}
\usepackage{verbatim}
\usepackage{tensor}
\usepackage{proof}
\usepackage[breaklinks,backref]{hyperref}
\usepackage{color}

\definecolor{aocolour}{rgb}{0.7,0.8,1}
\definecolor{mmcolour}{rgb}{1,0.8,0.7}

\newcommand{\set}[2]{\{ \, #1 \mid #2 \, \}}

\renewcommand{\emptyset}{\varnothing}
\renewcommand{\epsilon}{\varepsilon}
\renewcommand{\And}{\mathop{\&}}
\renewcommand{\tilde}{\widetilde}

\newcommand{\before}[1]{{\lhd} #1}
\newcommand{\beforeeq}[1]{{\trianglelefteqslant} #1}

\newcommand{\str}[2]{#1 \langle #2 \rangle}
\newcommand{\itm}[2]{#1\big(#2\big)}
\newcommand{\cps}[3]{\itm{#1}{\str{#2}{#3}}}

\newcommand{\odd}{\mathord{\textsc{Odd}}}
\newcommand{\even}{\mathord{\textsc{Even}}}
\newcommand{\uin}[3]{\tensor*[^{#1}]{#2}{^{#3}}}
\newcommand{\uinn}[4]{\tensor*[^{#1}]{#2}{^{#3}_{#4}}}
\newcommand{\lin}[3]{\tensor*[_{#1}]{#2}{_{#3}}}
\newcommand{\linn}[4]{\tensor*[_{#1}]{#2}{_{#3}^{#4}}}
\newcommand{\Eright}{\overrightarrow{E}}
\newcommand{\Erightzero}{\overrightarrow{E_0}}
\newcommand{\sublr}[3]{\lin{#1}{#2}{#3}}

\makeatletter
\newdimen\arrow@ht
\setbox\@tempboxa\hbox{\(\rightarrow\)}
\arrow@ht\ht\@tempboxa
\newdimen\star@wd
\setbox\@tempboxa\hbox{\(\scriptstyle *\)}
\star@wd\wd\@tempboxa
\def\rightstararrowfill@{\arrowfill@\relbar\relbar{\raisebox{0pt}[\arrow@ht][0pt]{\(\rightarrow^*\hskip-\star@wd\)}}}
\setbox\@tempboxa\hbox{\(\rightarrow\)}
\arrow@ht\ht\@tempboxa
\newcommand{\xrightstararrow}[2][]{\ext@arrow 0359\rightstararrowfill@{#1}{#2}\hskip\star@wd}
\makeatother

\newcommand{\rightarrowsymb}[1]{\xrightarrow{#1}}

\newtheorem{lemma}{Lemma}

\newtheorem{oldtheorem}{Theorem}

\numberwithin{remarktooldthm}{oldtheorem}
\newtheorem{claim}{Claim}
\numberwithin{claim}{lemma}

\numberwithin{oldclaim}{oldlemma}
\newtheorem{theorem}{Theorem}
\newtheorem{definition}{Definition}
\newtheorem{example}{Example}
\newtheorem{corollary}{Corollary}

\begin{document}

\sloppy

\title{The hardest language for grammars with context operators\thanks{%
	This work was supported by
	the Ministry of Science and Higher Education of the Russian Federation,
	agreement 075-15-2019-1619.
	}}
\author{Mikhail Mrykhin\thanks{%
	Department of Mathematics and Computer Science,
	St.~Petersburg State University, 7/9 Universitetskaya nab., Saint Petersburg 199034, Russia,
	\emph{and}
	Leonhard Euler International Mathematical Institute at St. Petersburg State University,
	Saint Petersburg, Russia.
	E-mail:
	\texttt{mikhail.k.mrykhin@gmail.com}.}
	\and
	Alexander Okhotin\thanks{%
	Department of Mathematics and Computer Science,
	St.~Petersburg State University, 7/9 Universitetskaya nab., Saint Petersburg 199034, Russia.
	E-mail:
	\texttt{alexander.okhotin@spbu.ru}.}
}

\maketitle

\begin{abstract}
In 1973, Greibach
(\href{http://dx.doi.org/10.1137/0202025}
	{``The hardest context-free language''},
	\emph{SIAM J. Comp.}, 1973)
constructed a context-free language $L_0$
with the property that every context-free language
can be reduced to $L_0$ by a homomorphism,
thus representing it as an inverse homomorphic image $h^{-1}(L_0)$.
In this paper, a similar characterization
is established for a family of grammars equipped with operators
for referring to the left context of any substring,
recently defined by Barash and Okhotin
(\href{http://dx.doi.org/10.1016/j.ic.2014.03.003}
	{``An extension of context-free grammars with one-sided context specifications''},
	\emph{Inform.\ Comput.}, 2014).
An essential step of the argument is a new normal form
for grammars with context operators,
in which every nonterminal symbol defines only strings of odd length
in left contexts of even length: the even-odd normal form.
The characterization is completed by showing that
the language family defined by grammars with context operators
is closed under inverse homomorphisms;
actually, it is closed under injective nondeterministic finite transductions.
\end{abstract}

\section{Introduction}

\emph{Grammars with context operators}
were defined by Barash and Okhotin~\cite{GrammarsWithContexts}
as an implementation of the vague idea of having a family of formal grammars
in which one could express a rule applicable only in contexts of a certain form.
Grammars with left context operators
generalize the ordinary formal grammars (Chomsky's ``context-free'');
they may use rules of the form
\begin{equation*}
	A \to BC \And \before{D},
\end{equation*}
which describe every substring representable as a concatenation $uv$,
with $u$ described by $B$ and $v$ described by $C$,
with the further condition that, to the left of $u$,
there is a substring of the form described by $D$.
In addition, grammars with left context operators
allow the conjunction of several syntactical constraints,
as in \emph{conjunctive grammars}~\cite{Conjunctive},
to be used freely;
one can use rules of the form 
\begin{equation*}
	A \to B_1 C_1 \And \ldots \And B_n C_n,
\end{equation*}
which describe all strings $w$
representable as each of the concatenations $B_i C_i$,
by some partition $w=u_i v_i$, with $u_i$ described by $B_i$ and $v_i$ described by $C_i$.

Being a further extension of conjunctive grammars,
grammars with left context operators
further improve their expressive power.
For instance, describing sequences of declarations and calls,
with the \emph{declaration before use} requirement,
is much easier than with conjunctive grammars~\cite{GrammarsWithContexts}.
Also grammars with left context operators can describe several
interesting abstract languages, such as $\set{ww}{w \in \{a,b\}^*}$~\cite{ConjunctiveTokyo}
and $\set{a^{n^2}}{n \geqslant 0}$~\cite{ContextsLinear}.

In spite of the increase in expressive power,
grammars with left context operators
still have efficient parsing algorithms.
Several algorithms are known.
The obvious algorithm runs in time $O(n^3)$~\cite{GrammarsWithContexts},
where $n$ is the length of the input string,
and its running time can be improved to $O(\frac{n^3}{\log n})$
by employing the Four Russians strategy~\cite{ContextsImproved}.
There is also a more practical variant of the Generalized LR,
with the running time between $O(n^4)$ and $O(n)$, depending on the grammar~\cite{ContextsLR}.
Also, there is a theoretical algorithm with space complexity $O(n)$~\cite{ContextsLinearSpace}.

Whether substantially subcubic-time parsing for these grammars is possible,
remains unknown.
Although parsing by matrix multiplication
extends to conjunctive grammars~\cite{BooleanMatrix},
these algorithms require reordering the computation steps
to the extent that make them inapplicable to grammars with contexts.

One of the classical results on the complexity of formal grammars
is Greibach's~\cite{Greibach} \emph{hardest context-free language},
which is an adaptation of the standard notion of a complete language in a complexity class to grammars,
using a \emph{homomorphism} as a reduction mechanism.
In other words, this hardest language $L_0$
allows every language $L$ defined by an (ordinary) grammar
to be represented as $L=h^{-1}(L_0)$, for a suitable homomorphism
(or as $h^{-1}(L_0 \cup \{\epsilon\})$, if $\epsilon \in L$).

For every family of languages, it is an interesting theoretical question
whether it has a hardest language under homomorphic reductions.
Already Greibach~\cite{Greibach_jump}
proved that the family of languages described by \emph{LR(1) grammars}
cannot have such a hardest language.
A similar negative result for the \emph{linear grammars}
was proved by Boasson and Nivat~\cite{BoassonNivat}.
On the other hand, Okhotin~\cite{ConjunctiveHardest}
has constructed the hardest language for conjunctive grammars,
whereas Mrykhin and Okhotin~\cite{MrykhinOkhotin_cellular} recently proved
that linear conjunctive grammars have no hardest language.
For the classical family of LL($k$) languages,
there is no hardest language in the strict sense,
that is, under homomorphic reductions~\cite{MrykhinOkhotin_LL};
however, if the reductions are relaxed to append a single end-marker to the homomorphic images,
then there is a single LL(1) language
which is hardest for the entire LL($k$) hierarchy~\cite{MrykhinOkhotin_LL}.

\begin{figure}[t]
	\centering
	\includegraphics[width=\textwidth]{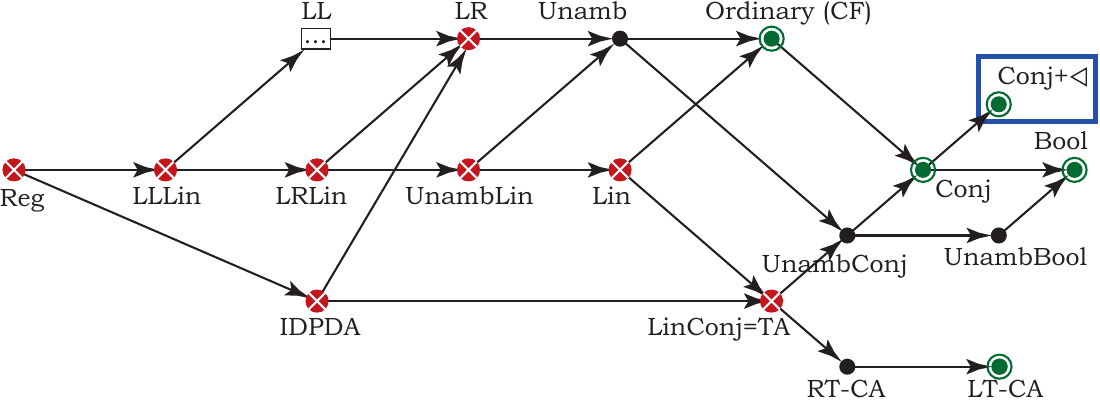}
	\caption{Existence (encircled) or non-existence (crossed out) of hardest languages
		in the hierarchy of formal languages:
		regular (Reg),
		ordinary grammars a.k.a.\ context-free (Ordinary) and their unambiguous subclass (Unamb),
		LL and LR grammars,
		input-driven a.k.a. visibly pushdown automata (IDPDA),
		linear grammars (Lin) and their subclasses (LLLin, LRLin, UnambLin),
		linear conjunctive grammars (LinConj),
		real-time and linear-time cellular automata (RT-CA, LT-CA),
		conjunctive grammars (Conj) and their unambiguous subclass (UnambConj).
		Boolean grammars (Bool) and their unambiguous subclass (UnambBool),
		grammars with left context operators ($\textrm{Conj}+\before{}$).
		For LL grammars, there is a hardest language only with an end-marker appended.}
	\label{f:hierarchy_hardest}
\end{figure}

Beyond formal grammars,
\v{C}ul\'{\i}k and Maurer~\cite{CulikMaurer} proved that there is no hardest regular language.
For one-counter automata, Autebert~\cite{Autebert_counter_invh}
also proved non-existence of hardest languages.
Mrykhin and Okhotin~\cite{MrykhinOkhotin_cellular} obtained the hardest language
for the family of linear-time cellular automata.
The results on hardest languages are illustrated in the hierarchy
presented in Figure~\ref{f:hierarchy_hardest}.

The goal of this paper is to construct the hardest language
for grammars with one-sided context operators.
Greibach's proof of her hardest language theorem for ordinary grammars
essentially uses the Greibach normal form,
with all rules of the form $A \to a \alpha$, where $a$ is a symbol of the alphabet.
Then, a homomorphic reduction of an arbitrary grammar to the hardest language
can assume a grammar in this normal form,
and use the image of the symbol $a$ to encode the entire grammar.

However, already for conjunctive grammars,
no analogue of the Greibach normal form is known,
and the construction of a hardest language relies on a more complicated
\emph{odd normal form}, established by Okhotin and Reitwie{\ss}ner~\cite{ConjunctiveNoUnion},
with all rules of the form $A \to a$ or $A \to B_1 a_1 C_1 \And \ldots \And B_n a_n C_n$,
where $a, a_1, \ldots, a_n$ are symbols of the alphabet,
and every nonterminal symbol defines only strings of odd length.
Then, the reduction to a hardest languages uses the images of $a_1, \ldots a_n$
to encode the grammar and to parse every conjunct $B_i a_i C_i$
from the image of $a_i$ outwards.

This paper begins with generalizing this normal form to grammars with context operators.
In Section~\ref{section_normal_form}, a new \emph{even-odd normal form} is introduced,
with the property that every substring defined in this grammar
is of odd length and is preceded by an even number of symbols
(that is, its left context is of even length).
A transformation to this normal form is presented.

Based on the even-odd normal form,
in Section~\ref{section_hardest},
a hardest language with respect to homomorphisms
for the family of grammars with one-sided context operators
is constructed.
The language is defined over a 6-symbol alphabet
and is given by a grammar with 14 nonterminal symbols and 35 rules.

A relevant question
is whether the language family defined by grammars with contexts
is closed under inverse homomorphisms.
As proved in Section~\ref{section_injective_transducers}, it is indeed closed:
in fact, closure under mapping implemented by injective nondeterministic finite transducers
is established.
This confirms that a language $L$ is defined by a grammar with left context operators
if and only if
it is representable as $L=h^{-1}(L_0)$, for some homomorphism $h$
(or as $h^{-1}(L_0 \cup \{\epsilon\})$, if $\epsilon \in L$).

\section{Grammars with one-sided context operators}\label{section_grammars}

For every partition of a string $w$ as $w=xyz$,
the string $y$ is a \emph{substring} of $w$,
the prefix $x$ is the \emph{left context} of $y$,
whereas the concatenation $xy$ is the \emph{extended left context} of $y$.
A substring $y$ written in a left context $x$
shall be denoted by $\str{x}{y}$ throughout this paper.

The family of \emph{grammars with left context operators}
allows a rule of the grammar
to define the properties of a substring
based not only on the structure of that substring,
but also on the structure of its left context
and its extended left context.

\begin{definition}[Barash and Okhotin~\cite{GrammarsWithContexts}]
\label{grammars_with_left_contexts_definition}
A grammar with left contexts is a quadruple $G=(\Sigma, N, R, S)$
that consists of the following components.
\begin{itemize}
\item
	A finite set of symbols $\Sigma$
	is the alphabet of the language being defined.
	Elements of $\Sigma$ are typically denoted
	by lower-case Latin letters from the beginning of the alphabet
	($a, b, \ldots$).
\item
	Another finite set $N$, disjoint with $\Sigma$,
	contains symbols
	for the syntactic properties of strings defined in the grammar
	(``nonterminal symbols'' in Chomsky's terminology).
	Symbols in $N$ are usually denoted by capital Latin letters.
\item
	A finite set of grammar rules $R$
	contains rules of the form
	\begin{equation}
	\label{eq:rule}
		A \to \alpha_1 \And \ldots \And \alpha_k \And
			\before{\beta_1} \And \ldots \And \before{\beta_m} \And
			\beforeeq{\gamma_1} \And \ldots \And \beforeeq{\gamma_n},
	\end{equation}
	where $A \in N$, $k \geqslant 1$, $m,n \geqslant 0$
	and $\alpha_i, \beta_i, \gamma_i \in (\Sigma \cup N)^*$.
	Informally, such a rule asserts that every substring
	representable as each concatenation $\alpha_i$,
	written in a left context representable as each $\beta_i$
	and in an extended left context representable as each $\gamma_i$,
	therefore has the property $A$.
\item
	The symbol $S \in N$
	represents the syntactically well-formed sentences of the language.
\end{itemize}
The size of $G$, denoted by $|G|$,
is the total number of symbols
used in the description of the grammar.
\end{definition}

A formal definition uses logical inference
on propositions of the form  $\cps{X}{u}{v}$,
with $X \in \Sigma \cup N$ and $u,v \in \Sigma^*$,
which means that
\emph{``a substring $v$
in the left context $u$
has the property $X$''}.

\begin{definition}[Barash and Okhotin~\cite{GrammarsWithContexts}]
Let $G=(\Sigma, N, R, S)$ be a grammar with left contexts,
and define the following deduction system
of elementary propositions of the form $\cps{X}{u}{v}$.
There is a single axiom scheme,
which asserts that a one-symbol substring $a \in \Sigma$
has the property $a$ in any left context $x \in \Sigma$.
\begin{align*}
	\infer{\cps{a}{x}{a}}{
	}
		&& (\text{for all $a \in \Sigma$ and $x \in \Sigma^*$})
\end{align*}
Each rule
(\ref{eq:rule})
in the grammar
defines a scheme for inference rules,
\begin{equation*}
	\infer{\cps{A}{u}{v}}{%
		I
	}
\end{equation*}
for all $u, v \in \Sigma^*$
and for every set of propositions $I$ satisfying the below properties:
\begin{enumerate}\renewcommand{\theenumi}{\roman{enumi}}
\item
	for every conjunct $\alpha_i=X_1 \ldots X_\ell$,
	with $\ell \geqslant 0$ and $X_j \in \Sigma \cup N$,
	there should exist a partition $v=v_1 \ldots v_\ell$,
	with $\cps{X_j}{u v_1 \ldots v_{j-1}}{v_j} \in I$
	for all $j \in \{1, \ldots, \ell\}$;
\item
	for every conjunct $\before{\beta_i}=\before{X_1 \ldots X_\ell}$,
	with $\ell \geqslant 0$ and $X_j \in \Sigma \cup N$,
	there should be such a partition $u=u_1 \ldots u_\ell$,
	that $\cps{X_j}{u_1 \ldots u_{j-1}}{u_j} \in I$
	for all $j \in \{1, \ldots, \ell\}$;
\item
	every conjunct $\beforeeq{\gamma_i}=\beforeeq{X_1 \ldots X_\ell}$,
	with $\ell \geqslant 0$ and $X_j \in \Sigma \cup N$
	should have a corresponding partition $uv=w_1 \ldots w_\ell$,
	with $\cps{X_j}{w_1 \ldots w_{j-1}}{w_j} \in I$ for all $j$.
\end{enumerate}
The condition in each case also applies if $\ell=0$
(that is, for conjuncts $\epsilon$, $\before{\epsilon}$ and $\beforeeq{\epsilon}$):
it degenerates
to $v=\epsilon$ for $\alpha_i=\epsilon$,
to $u=\epsilon$ for $\before{\beta_i}=\before{\epsilon}$,
and to $uv=\epsilon$ for $\beforeeq{\gamma_i}=\beforeeq{\epsilon}$.

A derivation of a proposition $\cps{A}{u}{v}$
is a sequence of such axioms and deductions,
where the set of premises at every step
consists of earlier derived propositions.
\begin{align*}
	I_1 &\vdash_G \cps{X_1}{u_1}{v_1} \\
	&\vdots \\
	I_{z-1} &\vdash_G \cps{X_{z-1}}{u_{z-1}}{v_{z-1}} \\
	I_z &\vdash_G \cps{A}{u}{v} \\
	& (\text{with }
		I_j \subseteq \set{\cps{X_i}{u_i}{v_i}}{i \in \{1, \ldots, j-1\}},
		\text{ for all } j)
\end{align*}
The existence of such a derivation
is denoted by $\vdash_G \cps{A}{u}{v}$.

Thus, for each symbol $A \in N$,
the following strings in contexts
have the property $A$.
\begin{equation*}
L_G(A) = \set{\str{u}{v}}{u,v \in \Sigma^*, \: \vdash_G \cps{A}{u}{v}}
\end{equation*}
The language described by the grammar $G$
is the set of all strings in left context $\epsilon$
that have the property $S$.
\begin{equation*}
L(G)=\set{w}{w \in \Sigma^{*}, \: \vdash_G \cps{S}{\epsilon}{w}}
\end{equation*}
\end{definition}

For more details on the definition,
the reader is referred to the original paper by Barash and Okhotin~\cite{GrammarsWithContexts},
as well as to a later paper by Okhotin~\cite{ContextsImproved}.

This definition is illustrated
on the following trivial example of a grammar.

\begin{example}\label{ab_trivial_example}
The following grammar with left contexts $G=(\Sigma, N, R, S)$
defines a single string $ab$.
\begin{equation*}
	\begin{array}{rcl}
		S &\to& AB \\
		A &\to& a \ | \ b \\
		B &\to& b \And \before{C} \\
		C &\to& a
	\end{array}
\end{equation*}
Without the context operator $\before{C}$,
the grammar would also define the string $bb$.
However, this context specification ensures that the first symbol must be $a$.

The string $ab$ is formally derived as follows.
\begin{equation*}
	\infer{\cps{S}{\epsilon}{ab}}
	{
		\cps{A}{\epsilon}{a}
		&
		\infer{\cps{B}{a}{b}}
		{
			\cps{C}{\epsilon}{a}
		}
	}
\end{equation*}

Note that the derivation
of $\cps{S}{\epsilon}{ab}$,
deriving the proposition $\cps{B}{a}{b}$
requires a left context of the form $C$.
The concatenation of $\cps{A}{\epsilon}{a}$ and $\cps{B}{a}{b}$
needed to infer $S$ respects contexts.
\end{example}

Among the basic properties of grammars with contexts
presented by Barash and Okhotin~\cite{GrammarsWithContexts},
there is a representation of derivations by parse trees,
and the following generalization of the Chomsky normal form.

\begin{oldtheorem}[Okhotin~\cite{ContextsImproved}]\label{binary_normal_form_theorem}
For every grammar with left contexts $G_0$,
there exists and can be effectively constructed
a grammar with left contexts $G=(\Sigma, N, R, S)$
that describes the language $L(G) = L(G_0) \setminus \{\epsilon\}$,
in which all rules in $R$ are of the following form.
\begin{subequations}\begin{align*}
	A &\to B_1 C_1 \And \ldots \And B_n C_n 
		&& (n \geqslant 1, \: B_i, C_i \in N)
	\\
	A &\to a \And \before{D}
		&& (a \in \Sigma, \: D \in N)
	\\
	A &\to a \And \before{\epsilon}
		&& (a \in \Sigma)
\end{align*}\end{subequations}

The size of $G$ is at most quadruple exponential in the size of $G_0$.
\end{oldtheorem}

The first step towards a hardest language theorem
is a new normal form presented in the next section.

\section{The even-odd normal form}\label{section_normal_form}

For conjunctive grammars, there is a normal form
known as the \emph{odd normal form}~\cite{ConjunctiveNoUnion},
in which all nonterminal symbols, except maybe the initial symbol,
define only strings of odd length.
In the following generalization of that normal form
to the case of grammars with contexts,
each nonterminal symbol defines strings of the form $\str{u}{v}$,
where the length of $v$ is odd and the length of its context $u$ is even.
The proposed normal form is accordingly called the \emph{even-odd normal form}.

\begin{definition}
A grammar with left contexts $G=(\Sigma, N, R, S)$ is in the even-odd normal form
if $S$ does not occur on the right-hand sides of any rules, and all its rules are of the following form.
\begin{align*}
	A &\to B_1a_1C_1 \And \ldots \And B_na_nC_n
		&& (B_i, C_i \in N, \: a_i \in \Sigma)
		\\
	A &\to a \And \before{Db}
		&& (D \in N, \: a, b \in \Sigma)
		\\
	A &\to a \And \before{\epsilon}
		&& (a \in \Sigma)
		\\
	S &\to Aa
		&& (A \in N, \: a \in \Sigma)
		\\
	S &\to \epsilon
\end{align*}
Furthermore, the rules of the last two forms are called even rules, and in their absence $G$ is said to be in the strict even-odd normal form.
\end{definition}

Let $\even$ be the set of all strings of even length over an implied alphabet $\Sigma$,
let $\odd$ similarly denote all strings of odd length.
Then the even-odd normal form clearly ensures that
$L_G(A) \subseteq \str{\Sigma^*}{\odd}$ for all $A \in N$
except maybe the initial symbol.
Upon a closer inspection,
one can see that 
$L_G(A) \subseteq \str{\even}{\odd}$,
whence the name of the normal form.

\begin{lemma}\label{even_odd_nf_string_parity_lemma}
Let $G=(\Sigma, N, R, S)$ be a grammar with left contexts in the even-odd normal form.
Then $L_G(A) \subseteq \str{\even}{\odd}$
for every nonterminal symbol $A \in N$
(except for $A=S$, if there are even rules for $S$).
\end{lemma}
\begin{proof}
It has to be proved that if, $\str{u}{v} \in L_G(A)$,
then $|u|$ is even and $|v|$ is odd.
The proof is by induction on the length of the proof of $\cps{A}{u}{v}$.

\begin{description}
\item[Base case: proof of length one, by a rule $A \to a \And \before{\epsilon}$.]
Then the proposition derived is $\cps{A}{\epsilon}{a}$,
where $\epsilon$ is of even length and $a$ is of odd length, as claimed.

\item[Induction step, rule $A \to a \And \before{Db}$.]
Assume that a proposition $\cps{A}{ub}{a}$ is derived using this rule.
\begin{equation*}
	\infer[(A \to a \And \before{Db})]{\cps{A}{ub}{a}}{
		\cps{D}{\epsilon}{u}
	}
\end{equation*}
Then it is derived from the premise $\cps{D}{\epsilon}{u}$,
which is accordingly derived in fewer steps than $\cps{A}{ub}{a}$.
Then, by the induction hypothesis, $|u|$ is odd,
and therefore $|ub|$ is even.

\item[Induction step, rule $A \to B_1a_1C_1 \And \ldots \And B_na_nC_n$.]
If $\cps{A}{x}{w}$ is derived using this rule,
then the last step of its derivation uses the following premises,
for some $n$ partitions of $w$ as $w=u_1 a_1 v_1= \ldots =u_n a_n v_n$.
\begin{equation*}
	\hspace*{-8mm}
	\infer[(A \to B_1a_1C_1 \And \ldots \And B_na_nC_n)]{\cps{A}{x}{w}}{
		\cps{B_1}{x}{u_1} & \cps{C_1}{xu_1a_1}{v_1}
		& \ldots
		& \cps{B_n}{x}{u_n} & \cps{C_n}{xu_na_n}{v_n}
	}
\end{equation*}
By the induction hypotheses for the derivations of these premises,
the length of $x$ is even,
whereas the lenghts of all $u_i$ and $v_i$ are odd.
Then $|w|$ is odd, as a sum of three odd numbers.
\qedhere
\end{description}
\end{proof}

The following theorem on the transformation to the even-odd normal form
shall now be proved.

\begin{theorem}
\label{even-odd_form_theorem}
For every grammar with left contexts $G=(\Sigma, N, R, S)$,
there exists a grammar with left contexts $G'=(\Sigma, N', R', S')$
in the even-odd normal form
that describes the same language.
The size of $G'$ is at most sextuple exponential in the size of $G$.
If $G$ is in the strong binary normal form,
then the blow-up is at most double exponential.
\end{theorem}

The resulting grammar $G'$ in the even-odd normal form
aims to recreate each parse tree in $G$.
The main difficulty is that the original parse of a string $w$ in $G$
may use propositions of the form $\cps{A}{u}{v}$,
without any restrictions on the parity of $|u|$ and $|v|$.
On the other hand,
when the length of $u$ is odd \emph{or} the length of $v$ is even,
according to Lemma~\ref{even_odd_nf_string_parity_lemma},
no grammar in the even-odd normal form
may define a node in a parse tree of $w$
spanning over this substring $v$.

The proposed solution is to simulate a node $A$
spanning over a substring from position $i$ to position $j$
with a node $A'$ spanning over a substring that begins in position $i$ or $i+1$
and ends in position $j$ or $j-1$.

\begin{figure}[t]
	\centerline{%
	\includegraphics[scale=1]{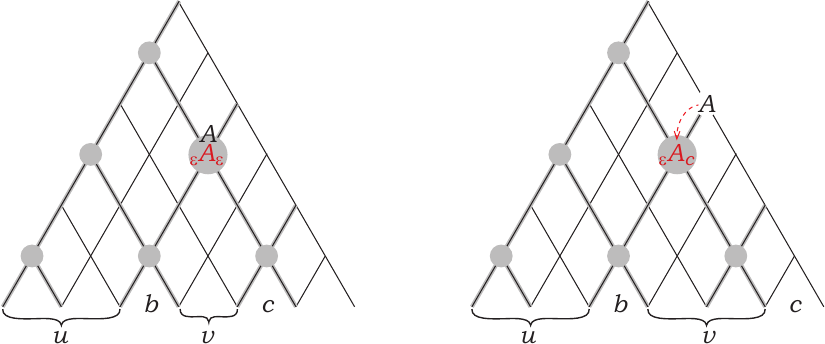}
	}
	\caption{(left) $|u|$ even, $|v|$ odd; (right) $|u|$ even, $|v|$ even.
		}
	\label{f:oddification_example_1}
\end{figure}

To be precise, let the original substring
be of the form $\str{u}{bvc}$, with $u, v \in \Sigma^*$ and $b, c \in \Sigma$.
If $|u|$ is even and $|bvc|$ is odd,
then the new grammar can have exactly the same node in its parse tree;
the corresponding nonterminal symbol in $G'$ shall be called $\sublr{\epsilon}{A}{\epsilon}$,
where empty strings on both sides indicate that
the substring in the new grammar
fits into exactly the same range of positions
as the substring in the original grammar.
This case is illustrated in Figure~\ref{f:oddification_example_1}(left),
in which grey circles indicate substrings of odd length with left contexts of even length,
and the string $\str{u}{bvc}$ falls into one of these grey circles.

If $|u|$ is even and $|bvc|$ is even,
then the new grammar shall define a substring $\str{u}{bv}$
by a nonterminal symbol $\sublr{\epsilon}{A}{c}$,
where $c$ indicates an outstanding symbol
that has to be appended in order to implement $A$.
This is shown in Figure~\ref{f:oddification_example_1}(right),
where the substring $\str{u}{bv}$ is the closest grey circle
to the original substring $\str{u}{bvc}$.

\begin{figure}[t]
	\centerline{%
	\includegraphics[scale=1]{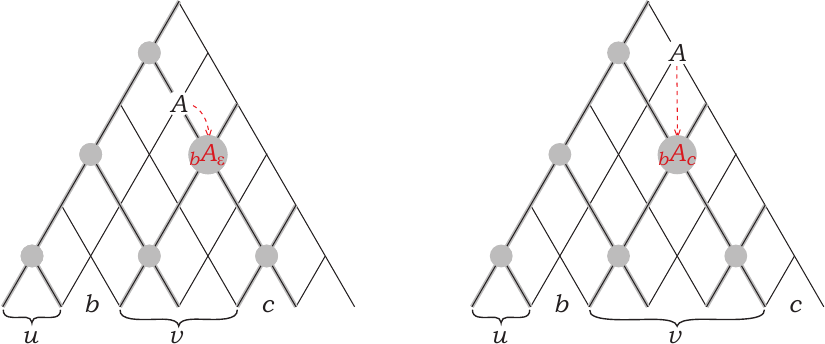}
	}
	\caption{(left) $|u|$ odd, $|v|$ even; (right) $|u|$ odd, $|v|$ odd.
		}
	\label{f:oddification_example_2}
\end{figure}

If $|u|$ is odd and $|bvc|$ is even,
then the corresponding string in the new grammar is $\str{ub}{vc}$, with $|ub|$ even and $|vc|$ odd,
defined by a nonterminal symbol called $\sublr{b}{A}{\epsilon}$,
with a symbol $b$ to be appended on the left.
This is the case in Figure~\ref{f:oddification_example_2}(left).
Finally, if $|u|$ is odd and $|bvc|$ is odd,
as illustrated in Figure~\ref{f:oddification_example_2}(right),
then the new grammar defines a string $\str{ub}{v}$, which has $|ub|$ even and $|v|$ odd,
by a nonterminal symbol $\sublr{b}{A}{c}$,
marking two symbols that need to be appended on both sides.

Overall, parse trees in $G'$
will generally reproduce the structure of original parse trees in $G$,
but some nodes shall be shifted by a couple of positions in the string.

The transformation to the even-odd normal form
shall be carried out in several stages.
Already at the first stage, given in Lemma~\ref{oddification_lemma} below,
the construction produces all nonterminal symbols that define only strings in $\str{\even}{\odd}$.
However, the languages defined are not exactly the same as in the original grammar,
and the rules may have conjuncts of several extra types.
In the rest of the transformations, conjuncts of unwanted types shall be gradually removed,
and at the final stage, the exact desired language shall be represented.

\begin{lemma}
\label{oddification_lemma}
For every grammar with left contexts $G$
there exists a grammar with left contexts $G_1 = (\Sigma, N_1, R_1, \lin{\epsilon}{S}{\epsilon})$
with $N_1 = (\Sigma \cup \epsilon) \times N \times (\Sigma \cup \epsilon)$,
where each nonterminal symbol $(x, A, y)$, denoted by $\lin{x}{A}{y}$ for convenience,
defines the following language.
\begin{equation*}
	L_{G_1}(\lin{x}{A}{y})
		=
	\set{\str{ux}{v}}{\str{u}{xvy} \in L_G(A), \str{ux}{v} \in \str{\even}{\odd}}
\end{equation*}
Furthermore, all conjuncts in $R_1$
are of the form $a$, $B$, $BaC$, $\beforeeq B$, $\before Ba$ or $\before\epsilon$,
with $B, C \in N$ and $a \in \Sigma$,
and each rule containing a conjunct $\before{\epsilon}$ also contains a conjunct $a$,
and each rule containing a conjunct $a$ also contains $\before{Ba}$ or $\before{\epsilon}$.
\end{lemma}

Assume that $G$ is in the strong binary normal form.
Then, define the set of rules $R_1$ of the new grammar,
which consists of the following rules.
\begin{subequations}\begin{align}
	\label{oddification_lemma__short_rule_with_context}
	\lin{\epsilon}{A}{\epsilon} &\to a \& \before \lin{\epsilon}{D}{b} b: &A \to a \&\before D \in R, b \in \Sigma,\\
	\label{oddification_lemma__short_rule}
	\lin{\epsilon}{A}{\epsilon} &\to a \& \before \epsilon : &A \to a \&\before\epsilon \in R,\\
	\label{oddification_lemma__long_rule}
	\lin{x}{A}{y} &\to\linn{x}{\alpha}{y}{(1)} \& \ldots \& \linn{x}{\alpha}{y}{(n)}:
		&A \to B^{(1)}C^{(1)} \&\ldots\& B^{(n)}C^{(n)} \in R, \\
\notag &&\mkern-100mu
\linn{x}{\alpha}{y}{(i)} \in \set{\linn{x}{B}{a}{(i)}a\linn{\epsilon}{C}{y}{(i)}}{a \in \Sigma}
\cup \set{\linn{x}{B}{\epsilon}{(i)}a\linn{a}{C}{y}{(i)}}{a \in \Sigma}\cup \\
\notag &&\cup\set{\linn{x}{B}{\epsilon}{(i)} \& \beforeeq \lin{\epsilon}{D}{\epsilon}}{C^{(i)} \to y \& \before D \in R}\cup \\
\notag &&\cup\set{\linn{\epsilon}{C}{y}{(i)} \& \before \lin{\epsilon}{D}{\epsilon}x}{B^{(i)} \to x \& \before D \in R}.
\end{align}\end{subequations}

\begin{proof}

First, it is claimed that $\str{ux}{v} \in L_{G_1}(\lin{x}{A}{y})$
if and only if
$\str{u}{xvy} \in L_G(A)$ and $\str{ux}{v} \in \str{\even}{\odd}$.
The proofs are separate in each directions and use induction on the length of the respective derivations.

\textcircled{$\supseteq$}
Most nonempty strings with contexts
are representable as $\str{u}{xvy}$,
where $x, y \in \Sigma \cup \{\epsilon\}$ and $\str{ux}{v} \in \str{\even}{\odd}$.
Indeed, the parity of $|u|$ determines whether to move the first symbol of the string into the contexts,
and the parity of the extended context determines whether the last symbol should be cut.
The only exception are strings $\str{\odd}{\Sigma}$,
where two symbols cannot be cut from a one-symbol string.
For all other strings, the representation exists and is unique.
\begin{description}
\item[Induction base:]
	$A(\str{u}{xvy})$ is derived in a single step
	if and only if it is derived by a rule of the form $A \to a\&\before\epsilon$,
	that is, $v = a, u = x = y = \epsilon$.
	Then $\lin{\epsilon}{A}{\epsilon}(\str{\epsilon}{a})$
	is derived by a rule $\lin{\epsilon}{A}{\epsilon}\to a \&\before\epsilon$.
	
\item[Induction step:]
	If $A(\str{u}{xvy})$ is derived by a rule $A \to a\&\before D$,
	then $v = a, x = y = \epsilon$, $D(\str{\epsilon}{u})$,
	and $u \in \even \setminus \{\epsilon\}$. Therefore $u = wb$ for some $w \in \odd$ and $b \in \Sigma$.
	Then, by the induction hypothesis, $\lin{\epsilon}{D}{b}(\str{\epsilon}{w})$ must be derivable,
	and then $\lin{\epsilon}{A}{\epsilon}(\str{wb}{a})$
	is derived by a rule $\lin{\epsilon}{A}{\epsilon} \to a \& \before \lin{\epsilon}{D}{b}b$.
	
	If $A(\str{u}{xvy})$, with $\str{ux}{v} \in \str{\even}{\odd}$,
	is derived by a rule of the form $A \to B^{(1)}C^{(1)} \And \ldots \And B^{(n)}C^{(n)}$.
	Then it is claimed that the grammar $G_1$
	contains a rule of the form (\ref{oddification_lemma__long_rule}),
	for some choice of conjuncts $\linn{x}{\alpha}{y}{(i)}$,
	by which one can derive $\str{ux}{v}$.
	
	For every $i$-th conjunct of the original rule,
	there is a partition $v = s_it_i$,
	with $B^{(i)}(\str{u}{xs_i})$ and $C^{(i)}(\str{uxs_i}{t_iy})$.
	
	First consider the case when both $s_i$ and $t_i$ are non-empty.
	If $s_i \in \odd$ and $t_i = aw_i$, with $a\in\Sigma$,
	then $w_i \neq \epsilon$, because $|s_i a w_i|$ is odd and $|s_i|$ is odd,
	and hence $|w_i|$ is odd.
	Then the induction hypothesis applies, and it asserts that
	$\linn{x}{B}{\epsilon}{(i)}(\str{ux}{s_i})$ and $\linn{a}{C}{y}{(i)}(\str{uxs_ia}{w_i})$.
	Accordingly, the conjunct $\linn{x}{\alpha}{y}{(i)}$
	in the rule (\ref{oddification_lemma__long_rule})
	is chosen as $\linn{x}{B}{\epsilon}{(i)} a \linn{a}{C}{y}{(i)}$,
	and it defines the string $\str{ux}{v}$.
	
	Similarly, if $s_i,t_i \neq \epsilon$, $s_i \in \even$ and $s_i = w_ia$, for $a\in\Sigma$,
	then both $|w_i|$ and $|t_i|$ are odd, and, by the induction hypothesis,
	$\linn{x}{B}{a}{(i)}(\str{ux}{w_i})$ and $\linn{\epsilon}{C}{y}{(i)}(\str{uxw_ia}{t_i})$.
	The conjunct deriving $\str{ux}{v}$ is then chosen as
	$\linn{x}{\alpha}{y}{(i)}=\linn{x}{B}{a}{(i)} a \linn{\epsilon}{C}{y}{(i)}$.
	
	If $t_i = \epsilon$, then $y \neq \epsilon$ and $s_i = v \neq\epsilon$.
	Therefore, $C^{(i)}(\str{uxs_i}{t_iy}) = C^{(i)}(\str{uxv}{y})$
	should be derived by a rule $C^{(i)} \to y \&\before D$,
	and this requires $D(\str{\epsilon}{uxv})$.
	Also $|uxv|$ is odd, because $\str{ux}{v} \in \str{\even}{\odd}$,
	and the induction hypothesis then implies $\lin{\epsilon}{D}{\epsilon}(\str{\epsilon}{uxv})$.
	Since $v = s_i$, the proposition $B^{(i)}(\str{u}{xv})$ is derived as well,
	and then $\linn{x}{B}{\epsilon}{(i)}(\str{ux}{v})$ by the induction hypothesis.
	The string $\str{ux}{v}$ is then defined by two conjuncts,
	$\linn{x}{B}{\epsilon}{(i)} \And \beforeeq{\lin{\epsilon}{D}{\epsilon}}$.
	
	Finally, if $s_i = \epsilon$, then $x \neq \epsilon$ and thus $x \in \Sigma$.
	Then $ux \in \even$ implies that $u \in \odd$ and $u \neq \epsilon$.
	In this case, $B^{(i)}(\str{u}{xs_i}) = B^{(i)}(\str{u}{x})$
	must be derived by a rule $B^{(i)}\to x \&\before D$, with $D\in N$,
	and this requires $D(\str{\epsilon}{u})$.
	The induction hypothesis then implies $\lin{\epsilon}{D}{\epsilon}(\str{\epsilon}{u})$.
	On the other hand, if $s_i = \epsilon$, then $v = t_i$,
	and $C^{(i)}(\str{ux}{vy})$ is derived.
	Then, by the induction hypothesis, $\linn{\epsilon}{C}{y}{(i)}(\str{ux}{v})$.
	The string $\str{ux}{v}$ is then defined by two conjuncts,
	$\linn{\epsilon}{C}{y}{(i)} \And \before{\lin{\epsilon}{D}{\epsilon}x}$.

	Overall, a rule of the form (\ref{oddification_lemma__long_rule})
	that defines the string $\str{ux}{v}$ has been constructed.	
\end{description}

\textcircled{$\subseteq$}
Conversely, if $\str{ux}{v} \in L_{G_1}(\lin{x}{A}{y})$,
then it should be proved, inductively on the length of the derivation,
that $\str{u}{xvy} \in L_G(A)$
and $\str{ux}{v} \in \str{\even}{\odd}$.

\begin{description}
\item[Induction base:]
	$\lin{x}{A}{y}(\str{ux}{v})$ is derived in one step
	if and only if
	it is derived by a rule of the form
	$\lin{\epsilon}{A}{\epsilon}\to a\&\before\epsilon$.
	Therefore, $u=x=y = \epsilon$ and $v = a$,
	and then $A(\str{\epsilon}{a})$ is derived by the original rule $A \to a \And \before\epsilon$.

\item[Induction step, short rule:]
	if $\lin{x}{A}{y}(\str{ux}{v})$
	is derived by a rule of the form
	$\lin{\epsilon}{A}{\epsilon} \to a \& \before \lin{\epsilon}{D}{b}b$,
	then $x = y = \epsilon$, $v = a$, $u = wb$, 
	and $\lin{\epsilon}{D}{b}(\str{\epsilon}{w})$ holds.
	Then, by the induction hypothesis, we can derive $D(\str{\epsilon}{wb})$,
	and $|w|$ is odd.
	The proposition $A(\str{wb}{a})$
	is then derived by the original rule $A \to a \&\before D$,
	and $|wb|$ is even, while $|a|=1$ is odd.
	
\item[Induction step, long rule:]
	Now, if $\lin{x}{A}{y}(\str{ux}{v})$ is derived by a rule of the form
	$\lin{x}{A}{y} \to \linn{x}{\alpha}{y}{(1)} \And \ldots \And \linn{x}{\alpha}{y}{(n)}$,
	obtained from a rule $A \to B^{(1)}C^{(1)} \And \ldots \And B^{(n)}C^{(n)}$ in the original grammar.
	Then every $i$-th $\linn{x}{\alpha}{y}{(i)}$ is a conjunct or a pair of conjuncts
	that define the string $\str{ux}{v}$,
	and it is claimed that $B^{(i)}C^{(i)}$ defines $\str{u}{xvy}$.
	The proof is different for each of the four types of conjuncts.
	
	If $\linn{x}{\alpha}{y}{(i)} = \linn{x}{B}{a}{(i)}a\linn{\epsilon}{C}{y}{(i)}$,
	then let $v = sat$,
	where $\linn{x}{B}{a}{(i)}(\str{ux}{s})$
	and $\linn{\epsilon}{C}{y}{(i)}(\str{uxsa}{t})$.
	Then, by the induction hypothesis, $B^{(i)}(\str{u}{xsa})$ and $C^{(i)}(\str{uxsa}{ty})$ hold,
	and therefore the concatenation $B^{(i)} C^{(i)}$
	defines the string $\str{u}{xsaty}=\str{u}{xvy}$.

 	The case of $\linn{x}{\alpha}{y}{(i)} = \linn{x}{B}{\epsilon}{(i)}a\linn{a}{C}{y}{(i)}$
	is considered similarly.
	
	In the case of a pair of conjuncts
	$\linn{x}{\alpha}{y}{(i)} = \linn{x}{B}{\epsilon}{(i)} \& \beforeeq \lin{\epsilon}{D}{\epsilon}$,
	it is given that $\linn{x}{B}{\epsilon}{(i)}(\str{ux}{v})$,
	$\lin{\epsilon}{D}{\epsilon}(\str{\epsilon}{uxv})$
	and the original grammar contains the rule $C^{(i)} \to y \And \before{D}$.
	The induction hypothesis is applicable to each of the above propositions,
	and it follows that $B^{(i)}(\str{u}{xv})$ and $D(\str{\epsilon}{uxv})$ hold.
	Furthermore, $C^{(i)}(\str{uxv}{y})$ can be derived by the rule for $C$.
	Then the concatenation $B^{(i)} C^{(i)}$
	produces the desired string as $\str{u}{xvy} = \str{u}{xv} \cdot \str{uxv}{y}$.
	
	If $\linn{x}{\alpha}{y}{(i)} = \linn{\epsilon}{C}{y}{(i)} \& \before \lin{\epsilon}D{\epsilon} x$,
	then $\linn{\epsilon}{C}{y}{(i)}(\str{ux}{v})$,
	$\lin{\epsilon}D{\epsilon}(\str{\epsilon}{u})$
	and there is a rule $B^{(i)} \to x \And \before{D}$.
	By the induction hypothesis, $C^{(i)}(\str{ux}{vy})$ and $D(\str{\epsilon}{u})$ hold,
	and by the aforementioned rule, $B^{(i)}(\str{u}{x})$ can be derived.
	Therefore $\str{u}{x} \cdot \str{ux}{vy}$ is the desired partition
	of $\str{u}{xvy}$ as $B^{(i)} C^{(i)}$.
	
	Since $\str{u}{xvy}$ is defined by each of the conjuncts of the rule for $A$,
	it follows that $\str{u}{xvy} \in L_G(A)$.
\end{description}

It remains to prove that, for every $\lin{x}{A}{y} \in N_1$,
each string in $L_{G_1}(\lin{x}{A}{y})$
is in $\str{\Sigma^* x}{\Sigma*}$, that is, its left context ends with $x$.
This is again proved by induction on the length of a derivation in $G_1$.
\begin{description}
\item[Induction base:]
	If $\lin{x}{A}{y}(\str{u}{v})$ is derived by a rule 
	\eqref{oddification_lemma__short_rule_with_context}
	or~\eqref{oddification_lemma__short_rule},
	then $x=\epsilon$, and the claim trivially holds.
	
\item[Induction step, long rule:]
	Assume that $\lin{x}{A}{y}(\str{u}{v})$
	is derived by a rule
	$\lin{x}{A}{y} \to \linn{x}{\alpha}{y}{(1)} \And \ldots \And \linn{x}{\alpha}{y}{(n)}$,
	obtained from a rule $A \to B^{(1)}C^{(1)} \And \ldots \And B^{(n)}C^{(n)}$.
	It is sufficient to consider $B^{(1)}C^{(1)}$ only.
	
	If $\linn{x}{\alpha}{y}{(1)}$
	is of the form $\linn{\epsilon}{C}{y}{(i)} \& \before \lin{\epsilon}{D}{\epsilon}x$,
	then the second conjunct in the pair ensures that $u$ ends with $x$, as desired.
	
	Otherwise, let $\linn{x}{\alpha}{y}{(1)}$
	be of the form $\linn{x}{B}{a}{(i)}a\linn{\epsilon}{C}{y}{(i)}$,
	$\linn{x}{B}{\epsilon}{(i)}a\linn{a}{C}{y}{(i)}$
	or
	$\linn{x}{B}{\epsilon}{(i)} \& \beforeeq \lin{\epsilon}{D}{\epsilon}$.
	Then, in each case, a nonterminal $\linn{x}{B}{z}{(i)}$ must define some string
	$\str{u}{s}$, where $s$ is a prefix of $v$.
	By the induction hypothesis, $u$ ends with $x$.
\end{description}
\end{proof}

As a second step of the transformation,
the rules of the new grammar
are transformed without affecting the set of nonterminal symbols
and the languages they define.
The resulting grammar will have no conjuncts of the form $B$, with $B \in N$,
called \emph{unit conjuncts}.
To be precise, all possible rules are grouped into the following three cases.

\begin{lemma}
\label{cleanup_lemma}
Let $G_1 = (\Sigma, N_1, R_1, S_1)$ be a grammar with left contexts,
in which the rules are comprised of conjuncts of the form
$a$, $B$, $BaC$, $\beforeeq B$, $\before Ba$ or $\before\epsilon$,
where $B, C \in N, a \in \Sigma$,
and a rule with an empty context must contain a solitary symbol $a$,
whereas a rule with a solitary $a$ must contain a context.
Then there exists a grammar with left contexts $G_2 = (\Sigma, N_1, R_2, S_1)$,
in which $L_{G_2}(A) = L_{G_1}(A)$ for each $A$ in $N_1$,
and all rules in $R_2$ are of the following three forms.
\begin{subequations}\begin{align}
	\label{cleanup_lemma__short_rule}
	A &\to a \&\before\epsilon,&&\mkern-50mu a \in \Sigma
		\\
	\label{cleanup_lemma__not_so_short_rule}
	A &\to a \&\before D_1b \&\ldots\&\before D_lb\&\beforeeq E_1\&\ldots\&\beforeeq E_m,\\
\notag	&&\mkern-400mu D_i, E_j \in N_1, b\in \Sigma, l \geqslant 1, m \geqslant 0
		\\
	\label{cleanup_lemma__long_rule}
	A &\to B_1a_1C_1 \And \ldots \And B_ka_kC_k \And
		\before{D_1 b} \And \ldots \And \before{D_l b} \And
		\beforeeq E_1 \And \ldots \And \beforeeq E_m,\\
\notag	&&\mkern-400mu B_i, C_i, D_j E_j \in N_1, \: a_i, b \in \Sigma,
		k \geqslant 1, \: l, m \geqslant 0
\end{align}\end{subequations}
Furthermore, if $G_1$ contains no extended contexts,
and all rules with contexts (now including non-empty ones) contain solitary symbols, 
then the same holds for $G_2$.
\end{lemma}
\begin{proof}
Start with removing redundancies from rules of the form $A \to a\&\before\epsilon\&\ldots$
Since they are only used for parsing strings of the form $\str{\epsilon}{a}$,
they can be replaced with at most $|\Sigma| \cdot |N_1|$ rules of the form $A \to a\&\before\epsilon$.

Then remove all remaining single nonterminal conjuncts.
This can be done by replacing them with all existing rules 
for the corresponding nonterminal~\cite{GrammarsWithContexts}.
If this step results in rules of the form $A \to a\&\before\epsilon\&\ldots$ reappearing,
they can be simply deleted, as the rules from step 1 
(which do not change during step 2) already make them redundant.

Proceed with removing all contradictory rules.
Specifically, rules that include both conjuncts of the form $a$ and $BaC$~---
the strings defined by $B$ and $C$ must be non-empty,
therefore, these conjuncts contradict each other,
and the rule cannot be used in any derivation.
Repeat for the rules containing both $\before\epsilon$ and $\before Db$
and these containing $a$ and $b$ or $\before Xa$ and $\before Yb$ for $a \neq b$.

After all these steps are taken, all conjuncts in 
the grammar still fit the original form, since only the first step creates 
new ones, which are valid. Conjunct $\before\epsilon$
can only appear in rules of the form $A \to a \&\before\epsilon$,
since it is contradictory to $\before Ba$ and
inseparable from $a$, which is in turn contradictory to $BaC$,
and both in conjunction make $\beforeeq E$ redundant.
Conjuncts of the form $a$ with no $\before\epsilon$
must still have a strict context operator in the same rule,
but cannot share it with $BaC$.
Finally, conjuncts of the form $BaC$ cannot share a rule with $a$,
and therefore, with $\before\epsilon$.
Since every rule contains at least one conjunct with no context operators,
this proves that the resulting ruleset satisfies the required restrictions.

The further restricted form is proved similarly. Since the first step does not create 
extended context operators, they do not appear, and since it only creates 
strict contexts together with solitary symbols, they stay together.
\end{proof}

At the next step, all extended context operators are eliminated.

\begin{lemma}
\label{deextension_lemma}
For every grammar with left contexts $G_2 = (\Sigma, N_1, R_2, S_1)$,
with all rules of the form as constructed in Lemma~\ref{cleanup_lemma},
there exists such a grammar $G_4 = (\Sigma, N_4, R_4, S_4)$
in strict even-odd normal form
with such subset of nonterminals $\set{\tilde{X}}{X \in N_2} \subset N_4$
that $\str{\epsilon}{v}$ is in $L_{G_4}(\tilde{A})$
if and only if it lies in $L_{G_2}(A)$.
\end{lemma}
\begin{proof}
We shall begin with constructing an intermediate grammar 
$G_3 = (\Sigma, N_3, R_3, S_3)$.
Let $P = 2^{N_1} \Sigma \cup \{\{\epsilon\}\}$
and $N_3 = P \times N_1 \times 2^{N_1}$,
where $(X, A, Y)$ is denoted by $\uin{X}{A}{Y}$ for convenience.
Each $\uin{X}{A}{Y}$ should define all strings $\str{u}{v}$
with the left context $\str{\epsilon}{u}$ satisfying all conditions listed in $X$,
which satisfy $A$ under the condition
that their extended contexts $\str{\epsilon}{uv}$ satisfy all conditions listed in $Y$.
This condition is then re-checked either by the strict context of the right concatenant 
or, if the strict context of the string is empty, by itself as its own extended context.

These conditional propositions are derived using the set of rules $R_3$, defined below.
The first type of rules apply to the first symbol of a string,
and the empty left context is preserved in $X=\{\epsilon\}$.
\begin{subequations}\begin{align}
	\label{deextension_lemma__short_rule}
	\uin{\{\epsilon\}}{A}{^\emptyset} &\to a \And \before\epsilon:
	&&\mkern-135mu A \to a \&\before\epsilon \in R_2,
\intertext{%
Every rule $A \to a \And \before D_1b \And \ldots \And \before D_lb
	\And \beforeeq E_1 \And \ldots \And \beforeeq E_m$
from the original grammar is simulated as follows:
all proper left contexts $D_ib$, as well as possible additional contexts,
are checked and recorded in $X$;
all extended left contexts $E_j$ are not checked and their list is remembered in $Y$,
to be checked later.
}
	\label{deextension_lemma__context_rule}
	\uin{\{H_i\}_ib}{A}{\{E_j\}_j} &\to a \And \before{\uinn{\{\epsilon\}}{H}{^\emptyset}{1}} b
	\And \ldots \And
	\before{\uinn{\{\epsilon\}}{H}{^\emptyset}{n} b}:\\
\notag&&\mkern-450mu A \to a \And \before D_1b \And \ldots \And \before D_lb
	\And \beforeeq E_1 \And \ldots \And \beforeeq E_m \in R_2,\\
\notag&&\mkern-450mu\{D_1, \ldots, D_l\} \subseteq \{H_1, \ldots, H_n\},
\intertext{%
A rule $A \to B_1a_1C_1 \And \ldots\And B_ka_kC_k
	\And \before{D_1 b} \And \ldots \And \before{D_l b}
	\And \beforeeq E_1 \And \ldots \And \beforeeq E_m$ from the original grammar
is simulated in the new grammar by a rule without any context operators.
For every original conjunct $B_i a_i C_i$,
the new rule contains a conjunct
$\uinn{X}{B}{Y_i}{i} a_i \uinn{Y'_ia_i}{C}{Z_i}{i}$,
where the set $Y'_i$ must contain $Y_i$
in order to verify every conditional context in $Y_i$.
The set $X$ is the same in all conjuncts,
and it is inherited by the nonterminal symbol defined in this rule.
The conditions $Z_i$ are accummulated in the nonterminal symbol defined,
and augmented with all $E_j$ from the proper contexts.
}
	\label{deextension_lemma__concatenation_rule}
	\uin{X}{A}{\{E_j\}_j \cup \bigcup_i Z_i} &\to
	\uinn{X}{B}{Y_1}{1} a_1 \uinn{Y'_1a_1}{C}{Z_1}{1}
	\And \ldots \And
	\uinn{X}{B}{Y_k}{k} a_k \uinn{Y'_ka_k}{C}{Z_k}{k}:\\
\notag&&\mkern-450mu
	A \to B_1a_1C_1 \And \ldots\And B_ka_kC_k
	\And \before{D_1 b} \And \ldots \And \before{D_l b}
	\And \beforeeq E_1 \And \ldots \And \beforeeq E_m \in R_2,\\
\notag&&\mkern-450mu\{D_1 b, \ldots, D_l b\} \subseteq X,
	\: Y_ia_i\subseteq Y'_ia_i
\intertext{%
The last type of rules
applies to subtrings with the empty left context,
and it allows any conditional extended context $E$
to be verified using a conjunction operator,
without any context operators.
}
	\label{deextension_lemma__deextension_rule} 
	\uin{\{\epsilon\}}{A}{Y} &\to \uin{\{\epsilon\}}{A}{Y\cup \{E\}} \And \uin{\{\epsilon\}}{E}{\emptyset}
\end{align}\end{subequations}

\begin{claim}
\label{reext_claim}
	If $\str{u}{v}$ is in $L_{G_3}(\uin{X}{A}{Y})$
	and $\str{\epsilon}{uv}$ lies in $L_{G_2}(E)$ for all $E$ in $Y$,
	then $\str{u}{v}$ is in $L_{G_2}(A)$
	and $\str{\epsilon}{u}$ lies in $L_{G_2}(\alpha)$ for all $\alpha \in X$.
\end{claim}

We shall prove this by induction on the length of derivation of $\uin{X}{A}{Y}(\str{u}{v})$.

\begin{description}
\item[Induction base:]
	The rules that perform single-step derivations
	are those and only those of the form~\eqref{deextension_lemma__short_rule}.
	In this case $X = \{\epsilon\}, Y = \emptyset, v = a, u = \epsilon$,
	and $R_2$ contains the rule $A \to a \And \before\epsilon$,
	which can be used to derive $A(\str{\epsilon}{a})$.
\item[Induction step:]
	If $\uin{X}{A}{Y}(\str{u}{v})$ is derived by a rule of the 
	form~\eqref{deextension_lemma__context_rule},
	then $v = a, X = Hb, u = wb$ for some $b\in\Sigma$,
	and $\uinn{\{\epsilon\}}{H}{^\emptyset}{i}(\str{\epsilon}{w})$ is derivable for all $1 \leqslant i \leqslant n$.
	Therefore, by the induction hypothesis we have $H_i(\str{\epsilon}{w})$ for all $1 \leqslant i \leqslant n$.
	Since $\{D_j\}_j \subseteq \{H_i\}_i$ and we are given 
	$E(\str{\epsilon}{uv})$ for all $E$ in $Y$,
	this is enough to derive $A(\str{u}{v}$ by the original rule of the form~\eqref{cleanup_lemma__not_so_short_rule}.
	
	If $\uin{X}{A}{Y}(\str{u}{v})$ is derived by a rule of the 
	form~\eqref{deextension_lemma__concatenation_rule},
	consider the $i$-th conjunct.
	According to it, $v = s_ia_it_i$ with $\uinn{X}{B}{Y_i}{i}(\str{u}{s_i})$ and
	$\uinn{Y'_ia_i}{C}{Z_i}{i}(\str{us_ia_i}{t_i})$.
	Given that $Z_i \subseteq Y = \left(\{E_1, \ldots,E_m\}\cup\left(\bigcup_i Z_i\right)\right)$ 
	and $Y_i \subseteq Y'_i$, the induction hypothesis says that we can derive
	$C_i(\str{us_ia_i}{t_i})$ and $H(\str{\epsilon}{us_i})$ for all $H \in Y_i$,
	which, in turn, lets us use the induction hypothesis once more to derive
	$B_i(\str{u}{s_i})$ and $\alpha(\str{\epsilon}{u})$ for all $\alpha \in X$.
	Since $\{D_i\}_ib \subseteq X$, we can use the original rule
	of the form~\eqref{cleanup_lemma__long_rule}.
	
	It remains to consider rules of the 
	form~\eqref{deextension_lemma__deextension_rule}.
	Applying the induction hypothesis to 
	$\uin{\{\epsilon\}}{E}{\emptyset}(\str{u}{v})$ 
	yields $u = \epsilon$ and $E(\str{u}{v})$. 
	The latter allows us to use the induction hypothesis
	for $\uin{\{\epsilon\}}{A}{Y\cup\{E\}}(\str{u}{v})$
	to prove $A(\str{u}{v})$, as desired.
\end{description}
\begin{claim}
\label{deext_claim}
\begin{enumerate}
\item
	If $\str{u}{v}$ is in $L_{G_2}(A)$
	and for some $X \in P$, $\str{\epsilon}{u}$ lies in $L_{G_2}(\alpha)$ for all $\alpha \in X$,
	then there is such $X' \in P, X' \supseteq X$, and $Y \subseteq N_2$
	that $\str{u}{v}$ is in $L_{G_3}(\uin{X'}{A}{Y})$
	and $\str{\epsilon}{uv}$ lies in $L_{G_2}(E)$ for all $E \in Y$,
	and furthermore, each $E(\str{\epsilon}{uv})$ with $E \in Y$
	has a shorter minimal derivation than $A(\str{u}{v})$.
	\item
	If $\str{\epsilon}{v}$ is in $L_{G_2}(A)$,
	then it is in $L_{G_3}(\uin{\{\epsilon\}}{A}{\emptyset})$ as well.
\end{enumerate}
\end{claim}

Both assertions are proved together in a single inductive argument 
on the sum of derivation lengths of $A(\str{u}{v})$ and all 
$H(\str{\epsilon}{ub^{-1}}$ with $Hb \in X$ (for the second 
assertion, this is just the derivation length of $A(\str{\epsilon}{v})$).
\begin{description}
\item[Induction base:]
	\begin{description}
	\item[]
	\item[2:]
	The sum of derivation lengths is 1 if and only if
	$A(\str{\epsilon}{v})$ is derived by rule of the 
	form~\eqref{cleanup_lemma__short_rule}.
	Therefore, the rule~\eqref{deextension_lemma__short_rule}
	can be used to derive $\uin{\{\epsilon\}}{A}{\emptyset}(\str{\epsilon}{v})$.
	\item[1:]
	The sum of derivation lengths is 1 if and only if
	$A(\str{u}{v})$ is derived by rule of the 
	form~\eqref{cleanup_lemma__short_rule}.
	It follows that $X = \{\epsilon\}, u = \epsilon$, 
	and we can apply point 2 to get 
	$\uin{X'}{A}{Y}(\str{\epsilon}{v})$, where
	$X' = \{\epsilon\}$ and $Y = \emptyset$ 
	(and thus the condition on $E\in Y$ holds trivially).
	\end{description}
\item[Induction step:]
	\begin{description}
	\item[]
	\item[1:]
	If $A(\str{u}{v})$ is derived by a rule of the 
	form~\eqref{cleanup_lemma__not_so_short_rule},
	then $v = a, u = wb\neq \epsilon, 
	D_i(\str{\epsilon}{w})$ for all $1\leqslant i\leqslant l$,
	and $E_j(\str{\epsilon}{uv})$ for all $1\leqslant j\leqslant m$.
	The induction hypothesis then claims that
	$\uinn{\{\epsilon\}}{D}{\emptyset}{i}(\str{\epsilon}{w})$
	and $\uinn{\{\epsilon\}}{E}{\emptyset}{j}(\str{\epsilon}{uv})$ 
	for all $1\leqslant i\leqslant l$ and $1\leqslant j\leqslant m$.
	At the same time,  $H(\str{\epsilon}{w})$ for all $Hb \in X$ 
	together with the induction hypothesis
	implies $\uin{\{\epsilon\}}{H}{\emptyset}(\str{\epsilon}{w})$
	for all $Hb \in X$.
	It remains to apply the rule~\eqref{deextension_lemma__context_rule}
	with $\{H_i\}_ib = \{D_i\}_ib \cup X$.
	
	If $A(\str{u}{v})$ is derived by a rule of the form~\eqref{cleanup_lemma__long_rule}
	and $H(\str{\epsilon}{ub^{-1}})$ for all $Hb \in X$,
	then for all $1\leqslant i\leqslant k$ there is a partition 
	$v = s_ia_it_i$ such that $B_i(\str{u}{s_i})$ and $C_i(\str{us_ia_i}{t_i})$,
	for all $1\leqslant j\leqslant l$ there is a partition 
	$u = wb$ such that $D_j(\str{\epsilon}{w})$,
	and for all $1\leqslant j' \leqslant m$ it holds  
	that $E_{j'}(\str{\epsilon}{uv})$.
	Then by the induction hypothesis (applied to $B_i$) there are such 
	$X' \in P$ with $\{D_j\}_jb\cup X \subseteq X'$ and $Y_i\subseteq N_2$,
	that for all $1\leqslant i\leqslant k$ we can derive $\uinn{X'}{B}{Y_i}{i}(\str{u}{s_i})$
	and for all $E \in Y_i$ we can derive $E(\str{\epsilon}{us_i})$.
	This allows us to apply the induction hypothesis to $C_i$,
	yielding for all $1\leqslant i\leqslant k$ such $Y'_i \supseteq Y_ia_i$ and $Z_i$,
	that $\uinn{Y'_i}{C}{Z_i}{i}(\str{us_ia_i}{t_i})$,
	and for all $E \in Z_i$
	the proposition $E(\str{\epsilon}{uv})$ has a shorter minimal derivation
	than $C_i(\str{us_ia_i}{t_i})$.
	It follows that, for each $E \in \{E_{j'}\}_{j'}\cup\left(\bigcup_i Z_i\right)$,
	the proposition $E(\str{\epsilon}{uv})$ has a shorter 
	minimal derivation than $A(\str{u}{v})$.
	It remains to use the rule
	$\uin{X'}{A}{\{E_{j'}\}_{j'}\cup\left(\bigcup_i Z_i\right)} \to \uinn{X'}{B}{Y_i}{i}a_i\uinn{Y'_ia_i}{C}{Z_i}{i}$
	to obtain $\uin{X'}{A}{\{E_{j'}\}_{j'}\cup\left(\bigcup_i Z_i\right)}(\str{u}{v})$.
	
	\item[2:]
	Applying point 1 for $X = \{\epsilon\}$,
	we get $X' = \{\epsilon\}$ and such $Y \subseteq N_2$
	that $\uin{\{\epsilon\}}{A}{Y}(\str{\epsilon}{v})$
	and for all $E \in Y$ the proposition $E(\str{\epsilon}{v})$
	has a shorter minimal derivation than $A(\str{\epsilon}{v})$.
	Then by the induction hypothesis (which is applicable thanks to shorter derivation) 
	for all $E \in Y$ we have $\uin{\{\epsilon\}}{E}{\emptyset}(\str{\epsilon}{v})$,
	which is enough to get  $\uin{\{\epsilon\}}{A}{\emptyset}(\str{\epsilon}{v})$ 
	from $\uin{\{\epsilon\}}{A}{Y}(\str{\epsilon}{v})$ by applying rules of the 
	form~\eqref{deextension_lemma__deextension_rule} $|Y|$ times.
	\end{description}
\end{description}

By Claim~\ref{deext_claim} and Claim~\ref{reext_claim} together,
$\str{\epsilon}{v}$ is in $L_{G_2}(A)$
if and only if
it lies in $L_{G_3}(\uin{\{\epsilon\}}{A}{\emptyset})$.
However, $R_3$ now includes single-nonterminal conjuncts, 
along with allowing multiple context operators in the same rule.
The former can be removed by applying Lemma~\ref{cleanup_lemma} 
again (this time in restricted form), while the latter can be eliminated by 
a powerset construction on the set of nonterminals~\cite{ContextsImproved}.
This produces a grammar $G_4 = (\Sigma, N_4, R_4, S_4)$, where $N_4 = 2^{N_3}$, 
and it remains to set $\tilde{X}$ as an alias for $\{\uin{\{\epsilon\}}{X}{\emptyset}\}$ 
to finish the proof.
\end{proof}

\begin{proof}[Proof of Theorem~\ref{even-odd_form_theorem}.]
Transform $G$ according to
Lemmata~\ref{oddification_lemma}, \ref{cleanup_lemma} and~\ref{deextension_lemma},
in this order.
This yields a grammar $G_4=(\Sigma, N_4, R_4, S_4)$.
It remains to add a new initial symbol $S'$.

Let $G' = (\Sigma, N_4 \cup \{S'\}, R_4\cup R_F, S')$,
where $R_F$ consists of the following rules:
\begin{align*}
S' &\to \Phi
	&& (\tilde{\lin{\epsilon}{S}{\epsilon}} \to \Phi \in R_4) \\
S' &\to \tilde{\lin{\epsilon}{S}{a}} a
	&& (a\in\Sigma) \\
S' &\to \epsilon
	&& (S \to \epsilon \in R)
\end{align*}
We need to prove that $L(G') = L(G)$.
Since all nonterminals in $G_1, G_2$ and $G_4$ only define strings of odd length,
$S'(\str{\epsilon}{v})$ can only be derived by a rule of the first form if $|v|$ is odd,
by a rule of the second form if $|v|$ is even and nonzero,
or by the last rule if $v$ is empty.
Let us consider all three cases.

First case: $|v|$ is odd.
Then, by Lemma~\ref{oddification_lemma},
a string $v$ lies in $L(G)$ if and only if
$\str{\epsilon}{v}$ lies in $L_{G_1}(\lin{\epsilon}{S}{\epsilon})$,
which, by Lemma~\ref{cleanup_lemma},
equals $L_{G_2}(\lin{\epsilon}{S}{\epsilon})$.
By Lemma~\ref{deextension_lemma}, this is equivalent to
$\str{\epsilon}{v}$ lying in $L_{G_4}(\tilde{\lin{\epsilon}{S}{\epsilon}})$,
which is in turn equivalent to $\str{\epsilon}{v}$ lying in $L_{G_4}(S')$, 
which is the definition of $v$ lying in $L(G_4)$.

Second case: $|v|$ is even and nonzero.
Let $v = ua, a\in\Sigma$.
Then, by Lemma~\ref{oddification_lemma}, $v$ lies in $L(G)$ 
if and only if $\str{\epsilon}{u}$ lies in $L_{G_1}(\lin{\epsilon}{S}{a})$,
which, by Lemma~\ref{cleanup_lemma}, equals $L_{G_2}(\lin{\epsilon}{S}{a})$.
By Lemma~\ref{deextension_lemma}, this is equivalent to
$\str{\epsilon}{u}$ lying in $L_{G_4}(\tilde{\lin{\epsilon}{S}{a}})$,
which is in turn equivalent to $\str{\epsilon}{ua}$ lying in $L_{G_4}(S')$, 
which is the definition of $ua = v$ lying in $L(G_4)$.

Third case: $v = \epsilon$.
Since the rule $S' \to \epsilon$ exists in $G'$ if and only if 
the rule $S \to \epsilon$ exists in $G$, and no other rule in either grammar 
can parse empty strings, this case is trivial.
\end{proof}

\section{Hardest language with left contexts}\label{section_hardest}

\begin{theorem}
\label{hardest_language_theorem}
There exists such language $L_0$ over the alphabet $\Sigma_0 = \{a, b, c, d, e, \#\}$
that it is described by a grammar with left contexts,
and any other language $L$ described by a grammar with left contexts
can be represented as $h_L^{-1}(L_0)$,
for some homomorphism $h_L \colon \Sigma^* \to \Sigma_0^*$, assuming that $\epsilon \notin L$
(where $\Sigma$ is the alphabet of $L$).
If $\epsilon \in L$, then $L=h^{-1}(L_0 \cup \{\epsilon\})$.
\end{theorem}
\begin{proof}
Without the loss of generality, assume that $L$ is described by a grammar 
$G = (\Sigma, N, R, S)$ in even-odd normal form.
Let $\mathcal{C} = \{\alpha_0, \ldots, \alpha_{ | \mathcal{C} |}\}$,
where $\alpha_i \in\{\epsilon, \before\epsilon\}\cup\Sigma\cup N\Sigma\cup\before N\Sigma\cup N\Sigma N$ 
is an enumeration of all conjuncts occurring in $R$,
augmented with strings $\beta \in \{\epsilon\}\cup N\Sigma$
corresponding to every conjunct $\before{\beta}$ in the grammar.
Then, every rule in $R$ is in the form $A \to \alpha_{i_1}\&\ldots\&\alpha_{i_m}$.
Also let us fix $\alpha_0 = \epsilon$.
The following construction generalizes
the hardest language for conjunctive grammars, as constructed by Okhotin~\cite{ConjunctiveHardest}.
We shall utilize the property of even-odd normal form 
that each non-empty conjunct contains exactly one terminal symbol;
this will allow us to encode the rules of $G$ into the images of these symbols.
After this we shall model the parsing in $G$ ``half-step off'', 
working with conjuncts instead of individual nonterminals.
\begin{itemize}
\item Symbols $a$ are used to represent references to a conjunct $\alpha_i$ as $a^i$.
\item The symbol $c$ is used to represent conjunction. For an arbitrary $r = \alpha_{i_1}\&\ldots\&\alpha_{i_m}$
its left and right representations are respectively
\begin{equation*}
\lambda(r) = ca^{i_1}\ldots ca^{i_m},
\quad \text{and} \quad 
\rho(r) = a^{i_m}c\ldots a^{i_1}c.
\end{equation*}
\item Symbols $b$ are used to mark rules for expanding a conjunct $\alpha_i$ as $b^i$.
\item An expansion of a conjunct $a_k = BaC$ consists of a marker $b^k$ 
preceded by a left representation of a rule $r$ for $B$ 
and followed by a right representation of a rule $r'$ for $C$, forming the string
$\lambda(r)b^k\rho(r')$. 
For a conjunct $\alpha_k = Ba$, the expansion accordingly omits $\rho(r')$, 
taking the form $\lambda(r)b^k$.
Similarly, a conjunct $\alpha_k = a$ is expanded as $b^k$, dropping both rules.
The last case are the conjuncts of the special form 
$\alpha_k = \before \alpha_l$, which slightly alter this construction. 
To represent the left context operator, a symbol $e$ is inserted between 
the left representation and the marker, giving the expansion the form of $\lambda(\alpha_l)eb^k$.

\item The symbol $d$ is used to separate different expansions of the same conjunct 
according to all combinations of rules for its constituent nonterminals,
forming the definition of said conjunct.
\begin{equation*}
	\sigma(\alpha_k) = \begin{cases}
		\prod_{B \to r, C \to r'} \lambda(r)b^k\rho(r')d, & \alpha_k = BaC \\
		\prod_{B \to r} \lambda(r)b^kd, & \alpha_k = Ba\\
		\prod \lambda(\alpha_l)eb^kd, & \alpha_k = \before \alpha_l\\
		b^kd, & \alpha_k = a
		\end{cases}
\end{equation*}
\item Finally, the full image of a symbol consists of definitions of all conjuncts that include
this symbol. Additionally, it includes a separate block of rule representations for the start symbol $S$, 
and an end-marker $\#$ to separate images of different symbols in the string:
\begin{equation*}
	h_G(s) = d\underbrace{(\prod_{S \to r}\rho(r)d)}_{h'(s)}d\underbrace{(\prod_{\alpha_k \in \mathcal{C}, s\in \alpha_k}\sigma(\alpha_k))}_{h''(s)}\#
\end{equation*}
\end{itemize}
To parse a substring according to some conjunct,
it is searched for a marker $b^n$ that matches the rule's $a^n$,
and then recursively parsed downwards according to the neighbouring markers $a^{i}$ again.
The hardest grammar $G_0$ uses the set of 14 nonterminals
$N_0 = \{S_0, A, B, C, D, \overrightarrow{E}, \overrightarrow{E_+}, \overrightarrow{F},
\overleftarrow{E}, \overleftarrow{E_+}, \overleftarrow{F}, \overleftarrow{H},
\overrightarrow{E_0}, \overrightarrow{F_0}\}$, the purpose of which is explained below, 
along with the rules of the grammar.

The main parsing work is done by the nonterminals $E$ and $F$,
which come in two directions depending on the direction of parsing 
(i.e. the direction in which the parsed substring is located, relative 
to the symbol containing the encoding of the parsing rule).
\begin{align*}
	\overbrace{a^{i_1} c a^{i_2} c \ldots a^{i_m} c d x \#
	\cdot
	\overbrace{h(u)
	\cdot
	x' \; \lambda(r')}^{\overleftarrow{E}} b^{i_1}}^{\overrightarrow{F}} \overbrace{\rho(r'') \; x''
	\cdot
	h(v)}^{\overrightarrow{E}}
	\;\in\;
	L_{G_0}(\overrightarrow{E})
\end{align*}
The nonterminal $\overrightarrow{E}$ handles the case when a rule is encoded 
at the left end of the current substring, thus parsing to the right of the rule.
It works by invoking $\overrightarrow{F}$ to match $a^{i_1}$
for the first conjunct with $b^{i_1}$ somewhere within the substring, and 
another instance of $\overrightarrow{E}$ to handle the right rule in the 
found expansion of $\alpha_{i_1}$. At the same time, it skips $a^{i_1}$
and proceeds to the rest via conjunction with $\overrightarrow{E_+}$.
\begin{align*}
	\overrightarrow{E} &\to \overrightarrow{F}\overrightarrow{E}\&Ac\overrightarrow{E_+}\\
	\overrightarrow{E_+} &\to \overrightarrow{F}\overrightarrow{E}\&Ac\overrightarrow{E_+}\\
	A &\to Aa \ |\ a
\intertext{%
As mentioned above, $\overrightarrow{F}$ matches 
$a$ on the left of its substring with the same number of $b$ on the right,
and then skips the rest of the symbol containing the previous rule, 
proceeding to invoke $\overleftarrow{E}$ onto the left rule in the 
expansion of $\alpha_{i_1}$. With that, both sides of the conjunct 
have been expanded.
}
	\overrightarrow{F} &\to a\overrightarrow{F}b \ |\ acC\#\overleftarrow{E}b \\
	C &\to aC \ |\ bC \ |\ cC \ |\ dC \ |\ eC \ |\ \epsilon
\intertext{%
Once there is no conjuncts, $\overrightarrow{E}$ or $\overrightarrow{E_+}$
conclude their work. Here the difference between them becomes apparent. 
Using $\overrightarrow{E_+}$ means that there are no more conjuncts, 
so the rest of the string (possibly including other images) is skipped. 
Meanwhile using $\overrightarrow{E}$ means that there were no conjuncts 
in the rule to begin with, that is, the rule is $\epsilon$, so no more images 
(beyond the current one) are allowed in the substring.
}
	\overrightarrow{E} &\to dC\#\\
	\overrightarrow{E_+} &\to dC\#D\\
	D &\to C\#D \ |\ \epsilon
\intertext{%
The left variations of $E$ and $F$ are parsed similarly.
}
	\overleftarrow{E} &\to \overleftarrow{E}\overleftarrow{F}\&\overleftarrow{E_+}cA \ |\ Cd\\
	\overleftarrow{E_+} &\to \overleftarrow{E}\overleftarrow{F}\&\overleftarrow{E_+}cA \ |\ DCd \\
	\overleftarrow{F} &\to b\overleftarrow{F}a \ |\ b\overrightarrow{E}Cca
\intertext{%
Additionally, $\overleftarrow{E}$ uses a special rule with no right-sided counterpart:
}
	\overleftarrow{E} &\to Cd\overleftarrow{H}e
\intertext{%
It invokes the new nonterminal $\overleftarrow{H}$, which performs the role 
of the left context operator. As rules with context operators do not contain 
other nonterminals, recursion to $\overleftarrow{E_+}$ is unnecessary.
}
	\overbrace{\text{context} \langle xd\overbrace{\lambda(\alpha_i)}^{\overleftarrow{H}}}^{\overleftarrow{E}}
	e\rangle \;&\in\; L_{G_0}(\overleftarrow{E})
\intertext{%
Depending on whether the referenced context is empty or nonempty, 
different context operators are used (either including the reference or not).
}
	\overleftarrow{H} &\to cA\&\beforeeq\overleftarrow{E}\ |\ c\&\before\overleftarrow{E}
\intertext{%
Finally, the starting symbol $S_0$ skips over an arbitrary number of rules 
(but does not pass the $dd$ marker separating the starting rules from proper rules), 
then invokes $\overrightarrow{E_0}$ and $\overrightarrow{F_0}$ similarly to 
$\overrightarrow{E}$.
}
	S_0 &\to dBS_0 \ |\ \overrightarrow{F_0}\overrightarrow{E}\&Ac\overrightarrow{E_0} \\
	B &\to aB \ |\ cB \ |\ a \ |\ c
\intertext{%
Here $\overrightarrow{E_0}$ is a starting variation of $\overrightarrow{E_+}$, 
while $\overrightarrow{F_0}$ is a starting variation of $\overrightarrow{F}$. 
Their rules are mostly the same, with one important change: $\overrightarrow{F_0}$
now parses a reference not from outside of the substring, but from inside of the first image.
Accordingly, instead of skipping the included part of that image, 
the excluded part is recovered back into the parsed substring 
by another use of the context nonterminal $\overleftarrow{H}$.
}
	\overrightarrow{E_0} &\to \overrightarrow{F_0}\overrightarrow{E}\&Ac\overrightarrow{E_0} \ |\ dC\#D \\
	\overrightarrow{F_0} &\to a\overrightarrow{F_0}b \ |\ ac\overleftarrow{H}b
\end{align*}

The nonterminals are mostly analogous to the ones used for the conjunctive case~\cite{ConjunctiveHardest}
with the exception of the newly introduced $H$, which is used to parse context-dependent rules.
We shall now prove that this construction is correct.
\begin{lemma}
\label{code_correctness_lemma}
Let $G = (\Sigma, N, R, S)$ be a grammar in the even-odd normal form,
$h_G = h: \Sigma^* \to \Sigma_0^*$ be the homomorphism defined above
and $G_0 = (\Sigma_0, N_0, R_0, S_0)$ be the grammar defined above.
Then the following holds:
\begin{enumerate}
\item
	\label{code_correctness_lemma__Eright_empty}
	A string $\str{x}{dy\#h(v)}$,
	where $x \in \Sigma_0^*$, $y \in \{a, b, c, d, e\}^*$, $xdy\# = h(u)$,
	$u, v \in \Sigma^*$,
	lies in $L_{G_0}(\overrightarrow{E})$
	if and only if $v = \epsilon$.
	
	A string of this form is in $L_{G_0}(\overrightarrow{E_+})$
	for every $v \in \Sigma^*$.
	
\item
	\label{code_correctness_lemma__Eleft_empty}
	A string $\str{h(u)}{h(v)yd}$,
	where
	$u, v \in \Sigma^*$,
	$y \in \{a, b, c, d, e\}^*$,
	lies in $L_{G_0}(\overleftarrow{E})$
	if and only if $v = \epsilon$.
	
	A string of this form is in $L_{G_0}(\overleftarrow{E_+})$
	for every $v \in \Sigma^*$.

\item
	\label{code_correctness_lemma__Eright_nonempty}
	A string $\str{x}{a^{i_m}c\ldots a^{i_1}cdy\#h(v)}$,
	where $x \in \Sigma_0^*$,
	$m > 0$,
	$i_1, \ldots, i_m > 0$,
	$xa^{i_m}c\ldots a^{i_1}cdy\# = h(u)$,
	$u, v \in \Sigma^*$,
	$y \in \{a, b, c, d, e\}^*$,
	lies in $L_{G_0}(\overrightarrow{E})$
	if and only if $\str{u}{v}$ lies in $\bigcap^m_{j = 1} L_G(\alpha_{i_j})$.
	The same holds for $\overrightarrow{E_+}$.
	
\item
	\label{code_correctness_lemma__Eleft_nonempty}
	A string $\str{h(u)}{h(v)xdca^{i_1}\ldots ca^{i_m}}$,
	where
	$u, v \in \Sigma^*$,
	$x \in \{a, b, c, d, e\}^*$,
	$m > 0$,
	$i_1, \ldots, i_m > 0$,
	lies in $L_{G_0}(\overleftarrow{E})$
	if and only if $\str{u}{v}$ lies in $\bigcap^m_{j = 1}L_G(\alpha_{i_j})$.
	The same holds for $\overleftarrow{E_+}$.
	
\item
	\label{code_correctness_lemma__Eleft_context}
	A string $\str{h(u)}{h(v)xdca^l e}$, 
	where
	$u, v \in \Sigma^*$,
	$x \in \{a, b, c, d, e\}^*$,
	$l \geqslant 0$,
	lies in $L_{G_0}(\overleftarrow{E})$
	if and only if $\str{\epsilon}{uv}$ lies in $L_G(\alpha_l)$.

\item
	\label{code_correctness_lemma__the_whole_string}
	A string $\str{x}{a^{i_m}c\ldots a^{i_1}cdyh''(t)\#h(v)}$,
	where
	$x \in \{a, b, c, d\}^+dd\cup\{d\}$,
	$m \geqslant 0$, 
	$i_1, \ldots, i_m > 0$,
	$xa^{i_m}c\ldots a^{i_1}cdy = h'(t)d$,
	$t \in \Sigma$,
	$v \in \Sigma^*$,
	lies in $L_{G_0}(\overrightarrow{E_0})$
	if and only if $\str{\epsilon}{tv}$ lies in $\bigcap_{j = 1}^mL_G(\alpha_{i_j})$.
	
\item
	\label{code_correctness_lemma__the_whole_string_start}
	A string $h(tv)$, where $t \in \Sigma$ and $v \in \Sigma^*$,
	lies in $L(G_0)$ if and only if $tv$ lies in $L(G)$.
\end{enumerate}
\end{lemma}
\begin{proof}
It is easy to see that $A, B, C$ and $D$
describe $a^+, \{a, c\}^+, (\Sigma_0\setminus\{\#\})^*$
and $\Sigma_0^*\#\cup\{\epsilon\}$ (independent of context), respectively.
\begin{description}
\item[\ref{code_correctness_lemma__Eright_empty}:]
	$\Leftarrow$: If $v = \epsilon$, then $\str{x}{dy\#h(v)} = \str{x}{dy\#}$
	can be parsed by the rule $\overrightarrow{E} \to dC\#$.
	
	$\Rightarrow$: If $\str{x}{dy\#h(v)}$ can be parsed as $\overrightarrow{E}$, 
	it cannot be done by the rule 
	$\overrightarrow{E} \to \overrightarrow{F}\overrightarrow{E}\&Ac\overrightarrow{E_+}$,
	as $A$ is always nonempty and starts with $a$ (instead of $d$).
	The other rule, $\overrightarrow{E} \to dC\#$, 
	requires there to be exactly one $\#$ in $dy\#h(v)$, 
	which is only possible if $h(v)$ is empty.
	
	For $\overrightarrow{E_+}$,
	a string $\str{x}{dy\#h(v)}$ is obtained using the rule $\overrightarrow{E_+} \to dC\#D$.

\item[\ref{code_correctness_lemma__Eleft_empty}:]
	$\Leftarrow$: If $v = \epsilon$, then $\str{h(u)}{h(v)yd} = \str{h(u)}{yd}$
	can be parsed by the rule $\overleftarrow{E} \to Cd$.
	
	$\Rightarrow$: If $\str{h(u)}{h(v)yd}$ can be parsed as $\overleftarrow{E}$, 
	it cannot be done by the rule 
	$\overleftarrow{E} \to \overleftarrow{E}\overleftarrow{F}\&\overleftarrow{E_+}cA$,
	as $A$ is always nonempty and ends with $a$ (instead of $d$).
	The other rule, $\overrightarrow{E} \to Cd$, 
	requires there to be no $\#$ in $h(v)yd$, 
	which is only possible if $h(v)$ is empty.
	
	For $\overleftarrow{E_+}$,
	a string $\str{h(u)}{h(v)yd}$ is obtained using the rule $\overleftarrow{E_+} \to DCd$.

\begin{figure}[t]
	\centering
	\includegraphics[scale=.85]{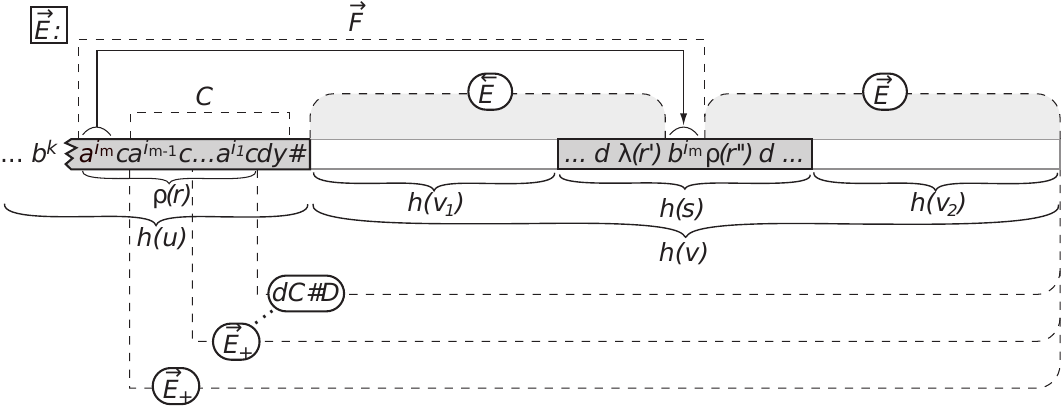}
	\caption{Case~\ref{code_correctness_lemma__Eright_nonempty}:
		the nonterminal $\protect\Eright$ parsing a string $\str{x}{\rho(r)dy\#h(v)}$
		in the case $\alpha_{i_m} = YsZ$.
		}
\end{figure}

\item[\ref{code_correctness_lemma__Eright_nonempty}:]
proved jointly with~\ref{code_correctness_lemma__Eleft_nonempty} and \ref{code_correctness_lemma__Eleft_context}, using induction on length of the string 
inside induction on the length of extended contexts (in other words, this is proved 
for any particular string after proving it for all other strings its parsing can depend on);
the proof of the reverse implication additionally uses induction on $m$.

Since $a^{i_m}c\ldots a^{i_1}cdy\#h(v)$ does not start with $d$,
the rule $\overrightarrow{E} \to dC\#$ is not applicable.
Consider the other rule
$\overrightarrow{E} \to \overrightarrow{F}\overrightarrow{E} \And Ac\overrightarrow{E_+}$.
The second conjunct is satisfied if and only if
$\str{xa^{i_m}c}{a^{i_{m-1}}c\ldots a^{i_1}cdy\#h(v)}$ lies in $L_{G_0}(\overrightarrow{E_+})$,
which by the induction hypothesis is equivalent to 
$\str{u}{v}$ lying in $\bigcap^{m-1}_{j = 1}L_G(\alpha_{i_j})$,
as long as $m \geqslant 2$.
If $m=1$, then $\str{u}{v}$ is trivially in $\bigcap^{m-1}_{j = 1}L_G(\alpha_{i_j})$,
whereas 
$\str{xa^{i_1}c}{dy\#h(v)}$ is in $L_{G_0}(\overrightarrow{E_+})$
by Case~\ref{code_correctness_lemma__Eright_empty}.

It remains to prove that the first conjunct $\overrightarrow{F}\overrightarrow{E}$ is satisfied
if and only if $\str{u}{v}$ is in $L_G(\alpha_{i_m})$.
By repeatedly expanding $\overrightarrow{F}$ in it,
we have that $\str{x}{a^{i_m}c\ldots a^{i_1}cdy\#h(v)}$
lies in $L_{G_0}(a^ncC\#\overleftarrow{E}b^n\overrightarrow{E})$ for some $n$.
Since $\overleftarrow{E}$ cannot end with $b$
while $\overrightarrow{E}$ cannot start with it,
and since $n = i_m$,
the substring $b^n$ in the partition above
is part of the expansion of $\alpha_{i_m}$
in the image of some symbol from $v$.
Consider the form of $\alpha_{i_m}$.

If $\alpha_{i_m} = YsZ$,
then $a^{i_m}c\ldots a^{i_1}cdy\#h(v)$
lies in $L_{G_0}(\overrightarrow{F}\overrightarrow{E})$ 
if and only if there is such a partition $v = v_1sv_2$ with $v_1, v_2\in\Sigma^*$ and $s\in\Sigma$,
two rules $Y \to r'$ and $Z \to r''$,
and such a partition $h(s) = y'd\lambda(r')b^{i_m}\rho(r'')dz'\#$,
that $\str{h(u)}{h(v_1)y'd\lambda(r')}$ lies in $L_{G_0}(\overleftarrow{E})$,
while $\str{h(u)h(v_1)y'd\lambda(r')b^{i_m}}{\rho(r'')dz'\#h(v_2)}$
lies in $L_{G_0}(\overrightarrow{E})$.
By the induction hypothesis this is equivalent to 
$\str{u}{v_1}$ lying in $L_G(\alpha)$ for all $\alpha \in r'$,
therefore, in $L_G(Y)$, and a similar argument holds for the right substring.

The cases $\alpha_{i_m} = Ys$ and $\alpha_{i_m} = s$ are considered similarly,
with the exception that whenever $b^{i_m}$ has no neighbouring symbol $a$ on the left or on the right,
the corresponding $v_j$ must be empty,
by the Cases~\ref{code_correctness_lemma__Eright_empty}--\ref{code_correctness_lemma__Eleft_empty}.

Finally, if $\alpha_{i_m} = \before{\alpha_l}$, then $v_2$ must again be empty 
(since, by the construction of context expansions, there is no rule on the right),
while $\str{h(u)}{h(v_1)xdca^l e}$ must be in $L_{G_0}(\overleftarrow{E})$. 
By the induction hypothesis (Case~\ref{code_correctness_lemma__Eleft_context}),
this is equivalent to $\str{\epsilon}{uv_1}$ lying in $L_G(\alpha_l)$, which is the same as 
$\str{uv_1}{s}$ lying in $L_G(\before\alpha_l)$. 
Note that we needed to prove this for $\str{u}{v}$; however, 
rules with contexts always start with a solitary terminal conjunct, 
therefore by the induction hypothesis $\str{u}{v} = \str{u}{s} = \str{uv_1}{s}$.

\begin{figure}[t]
	\centering
		\includegraphics[scale=.85]{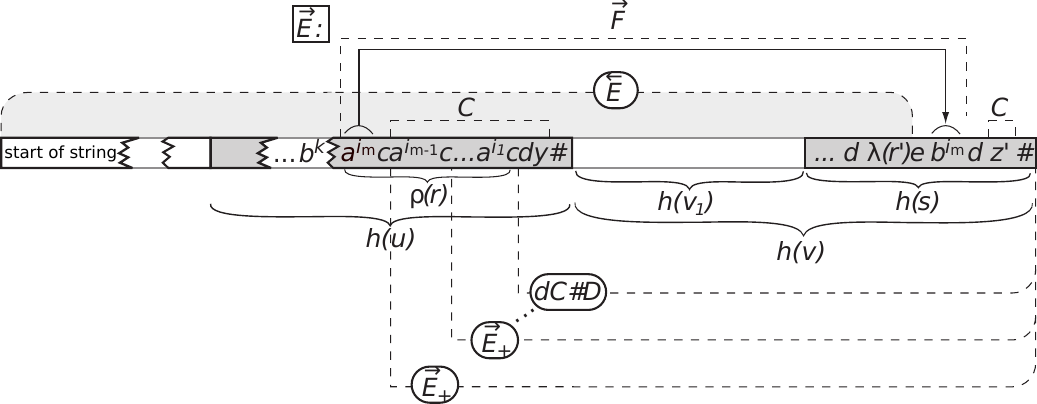}
	\caption{Case~\ref{code_correctness_lemma__Eright_nonempty}:
		the nonterminal $\protect\Eright$ parsing a string $\str{h(u)}{\rho(r)dy\#h(v)}$
		in the case of $\alpha_{i_m} = \before\alpha_j$.
		}
\end{figure}

\item[\ref{code_correctness_lemma__Eleft_nonempty}:] Similar to \ref{code_correctness_lemma__Eright_nonempty}.

\item[\ref{code_correctness_lemma__Eleft_context}:]
This is the case of a context operator $\before\alpha_l$.
Since the string $\str{h(u)}{h(v)xdca^l e}$ ends with an $e$,
the only rule for $\overleftarrow{E}$ applicable to it
is the rule $\overleftarrow{E} \to Cd\overleftarrow{H}e$.
By substituting the rule for $\overleftarrow{H}$, 
we end up with one of two options:
either $l = 0$ (i.e. $\alpha_l = \epsilon$)
and $\str{\epsilon}{h(uv)xd}$ lies in $L_{G_0}(\overleftarrow{E})$, 
which, by Case~\ref{code_correctness_lemma__Eleft_empty},
is equivalent to $uv = \epsilon$;
or $l > 0$, and $\str{\epsilon}{h(uv)xdca^l}$ lies in $L_{G_0}(\overleftarrow{E})$,
which, by Case~\ref{code_correctness_lemma__Eleft_nonempty},
is equivalent to $uv \in L_G(\alpha_l)$. 

\begin{figure}[t]
	\centering
		\includegraphics[scale=.85]{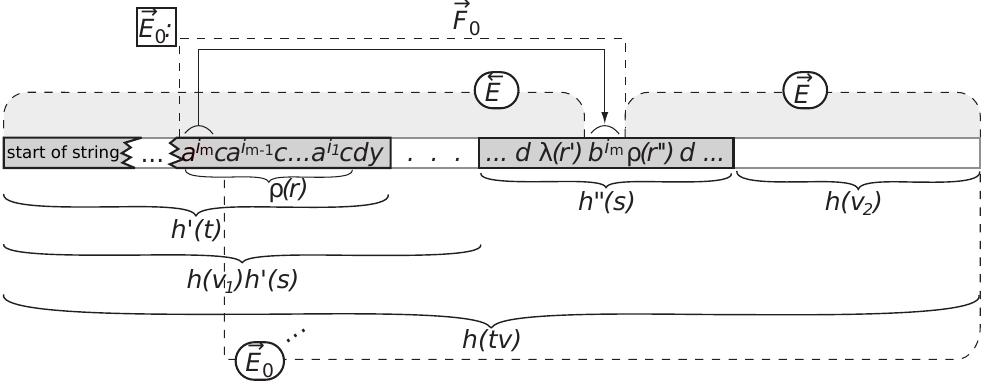}
	\caption{Case~\ref{code_correctness_lemma__the_whole_string}:
		the nonterminal $\protect\Erightzero$ parsing a string $\str{x}{\rho(r)dyh''(t)\#h(v)}$.
		}
\end{figure}

\item[\ref{code_correctness_lemma__the_whole_string}:]
Induction by $m$. The base $m=0$ is trivial by the rule $\overrightarrow{E_0} \to dC\#D$, 
as the string takes the form of $\str{x}{dyh''(t)\#h(v)}$.
The step is similar to Case~\ref{code_correctness_lemma__Eright_nonempty},
with a slight change in the rule for $\overrightarrow{F_0}$
searching for a partition of $tv$ instead of just $v$.

\item[\ref{code_correctness_lemma__the_whole_string_start}:]
$\Rightarrow:$ Induction by the length of the string.
If derivation uses the rule $S_0 \to dBS_0$, use the induction hypothesis 
for the shorter $S_0$.
Otherwise, use Case~\ref{code_correctness_lemma__the_whole_string}.

$\Leftarrow:$ Find the representation of the rule that parses
$\str{\epsilon}{tw}$ in the initial segment of image of $t$.
Skip the preceding representations with the rule $S_0 \to dBS_0$,
then use Case~\ref{code_correctness_lemma__the_whole_string}.
\end{description}
\end{proof}
The final claim of the final lemma effectively proves the theorem.
\end{proof}

\section{Closure under injective finite transductions}\label{section_injective_transducers}

It has been proved that every language defined by a grammar with left context operators
is representable as $h^{-1}(L_0)$ or as $h^{-1}(L_0 \cup \{\epsilon\})$,
for a single language $L_0$ defined by a grammar with left contexts.
It is natural to ask whether, conversely,
all inverse homomorphic images of $L_0$
are defined by grammars with left contexts,
that is, whether this family is closed under inverse homomorphisms.

The answer is positive:
in fact, similarly to unambiguous grammars~\cite{GinsburgUllian_preservation},
conjunctive grammars and Boolean grammars~\cite{BooleanInvGSM},
the family of grammars with left context operators
is closed under \emph{injective finite transductions},
and this can be proved by a straightforward generalization
of the classical construction.

\begin{definition}[Elgot and Mezei~\cite{ElgotMezei}]
A nondeterministic finite transducer (NFT)
is a sextuple $\mathcal{T}=(\Sigma, \Omega, Q, Q_0, \delta, F)$,
formed of the following components:
\begin{itemize}
\item
	a finite non-empty input alphabet $\Sigma$;
\item
	a finite non-empty output alphabet $\Omega$,
\item
	a finite non-empty set of states $Q$,
\item
	the set of initial states $Q_0 \subseteq Q$,
\item
	a finite set of transitions
	$\delta \subset Q \times (\Sigma\cup\{\epsilon\}) \times (\Omega\cup\{\epsilon\}) \times Q$,
	called the transition relation,
	and
\item
	the set of accepting states $F \subseteq Q$.
\end{itemize}
An NFT defines a multiple-valued function
$\mathcal{T} \colon \Sigma^* \to 2^{\Omega^*}$
that maps each string to a set of zero or more translations.

An computation of an NFT
is any sequence of transitions
passing through some states $p_0, p_1, \ldots, p_{\ell-1}, p_\ell \in Q$
while reading input strings $u_1, \ldots, u_\ell \in \Sigma$
and emitting output strings $x_1, \ldots, x_\ell \in \Omega$.
This sequence conforms to the transition function at each step,
as $(p_{i-1}, u_i, x_i, p_i) \in \delta$,
and is denoted as follows.
\begin{equation*}
	p_0
		\rightarrowsymb{u_1/x_1}
	p_1
		\rightarrowsymb{u_2/x_2}
	\ldots
		\rightarrowsymb{u_{\ell-1}/x_{\ell-1}}
	p_{\ell-1}
		\rightarrowsymb{u_\ell/x_\ell}
	p_{\ell}
\end{equation*}
Altogether, this is a computation from $p_0$ to $p_\ell$
that reads a string $u_1 \ldots u_\ell$
and emits a string $x_1 \ldots x_\ell$.
The existence of such a computation is represented by the following notation using a single arrow.
\begin{equation*}
	p_0
		\rightarrowsymb{u_1 \ldots u_\ell/x_1 \ldots x_\ell}
	p_{\ell}
\end{equation*}

Whenever there is a computation
from the initial state
to any accepting state
that reads a string $w \in \Sigma^*$
and emits a string $z \in \Omega^*$,
this sets $z$ as one of the possible translations of $w$.
Thus, the transducer defines the following translations
for each string $w \in \Sigma^*$.
\begin{equation*}
	\mathcal{T}(w)=\set{z}{q_0 \rightarrowsymb{w/z} q, \text{ for some } q \in F}
\end{equation*}
\end{definition}

An NFT $\mathcal{T}$ is called \emph{injective},
if, for every two distinct strings,
their images under $\mathcal{T}$ are disjoint.
Note, however, that injectivity does not necessarily 
imply that every string has at most one image.

\begin{theorem}
\label{injective_image_theorem}
Let $\mathcal{T}$ be an injective NFT mapping $\Sigma^*$ to $\Omega^*$.
Then, for every grammar with left context operators $G$ over the alphabet $\Sigma$,
there exists a grammar with left context operators $G'$ over the alphabet $\Omega$
that defines the language $L(G') = \mathcal{T}(L(G))$.
\end{theorem}

Let $G = (\Sigma, N, R, S)$ be the grammar,
and assume that it is in the strict binary normal form.
Let the transducer be $\mathcal{T} = (\Sigma, \Omega, Q, \{q_0\}, \{ q_F\}, \delta)$; 
the sets of initial and accepting states can be assumed to contain exactly one 
state each, since adding such states with $\epsilon$-transitions produces 
an equivalent transducer. 
Furthermore, we will also assume that every state is reachable from $q_0$, 
and that $q_F$ is reachable from every state 
(since the transition relation does not have to be total, ``dead'' states 
can be simply removed).
This will assure that every computation is uniquely identified by its starting state, 
ending state and output string; otherwise there would be two different strings with 
a common image, as both computations could be taken to proper accepting 
computations by adding the same prefix and suffix.

The new grammar $G'$ will contain nonterminals of two kinds.
First, there are nonterminals of the form $A_{p,q}$, where $p, q \in Q$ and $A \in N$;
this nonterminal should define all strings that the transducer can emit
while reading a string in $L_G(A)$, 
if it begins in the state $p$ and finishes reading it in the state $q$.
The other type of nonterminals are of the form $\epsilon_{p,q}$:
such a nonterminal defines all strings that the transducer can emit,
if it moves from $p$ to $q$ without reading any input symbols.
Finally, there is a start symbol $S_0$.

The pseudoempty nonterminals $\epsilon_{p,q}$ serve as the initialization 
step of the transformed grammar.
\begin{align*}
	\epsilon_{p,q} &\to b
	&& b \in \Omega\cup\{\epsilon\}, \: p \rightarrowsymb{\epsilon/b} q
\intertext{%
As the original transition relation both consumes and produces at most one symbol 
at each computation step, strings produced by consuming nothing 
can be easily constructed from separate symbols.
}
	\epsilon_{p,q} &\to \epsilon_{p,r}\epsilon_{r,q}
\intertext{%
Next to be transformed are the single-terminal strings. Since the construction has 
to describe all computations, they are additionally 
padded with images of the empty string.
}
	A_{p,q} &\to \epsilon_{p,r}b\epsilon_{r',q} \& \before \epsilon_{q_0,p}
	&& A \to a \& \before \epsilon \in R, \: b \in \Omega\cup\{\epsilon\}, \: r \rightarrowsymb{a/b} r'
		\\
	A_{p,q} &\to \epsilon_{p,r}b\epsilon_{r',q} \& \before D_{q_0,p}
	&& A \to a \& \before D \in R, \: b \in \Omega\cup\{\epsilon\}, \: r \rightarrowsymb{a/b} r'
\intertext{%
Then the concatenation rules are transformed. This is where injectivity is vital: 
without it, there would be no guarantee that the preimages of $B^{(1)}C^{(1)}$ 
are the same string for all $i$, and so the rule would have a possibility 
of defining strings not in the image of $L_G(A)$. 
Also note that all constituent nonterminals are already 
padded with empty images, so no more are required.
}
	A_{p,q} &\to B^{(1)}_{p,r_1}C^{(1)}_{r_1,q}\&\ldots\& B^{(n)}_{p,r_n}C^{(n)}_{r_n,q}
	&& A \to B^{(1)}C^{(1)}\&\ldots \&B^{(n)}C^{(n)} \in R,
	r_1, \ldots, r_n \in Q
\intertext{%
Finally, we have the starting rules, which are effectively just an alias.
}
	S_0 &\to S_{q_0, q_F}
		\\
	S_0 &\to  \epsilon_{q_0, q_F}
	&& \text{only if $S \to \epsilon$}
\end{align*}

\begin{lemma}
\label{epsilon_image_lemma}
Each nonterminal symbol $\epsilon_{p,q}$ in the constructed grammar,
with $p, q \in Q$,
defines the language
$L_{G'}(\epsilon_{p,q}) = \set{\str{\Omega^*}{z}}{p \rightarrowsymb{\epsilon/z} q}$.
\end{lemma}
\begin{proof}
\textcircled{$\subseteq$}: proof by induction on the length of derivation.

Induction base: if $\epsilon_{p,q}(\str{x}{y})$ is derived in one step, 
it must use a rule of the form $\epsilon_{p,q} \to b$. 
Then $b = y$, and by the construction of the rules, 
$p \rightarrowsymb{\epsilon/y} q$.

Induction step: if $\epsilon_{p,q}(\str{x}{y})$ is derived in more than one step, 
it must use a rule of the form $\epsilon_{p,q} \to \epsilon_{p,r}\epsilon_{r,q}$. 
Then $y = y_1y_2$, with $\str{x}{y_1} \in L_{G'}(\epsilon_{p,r})$ and 
$\str{xy_1}{y_2} \in L_{G'}(\epsilon_{r,q})$. By the induction hypothesis
we have $p \rightarrowsymb{\epsilon/y_1} r$ and $r \rightarrowsymb{\epsilon/y_2} q$, 
which are easily composed into $p \rightarrowsymb{\epsilon/y_1y_2} q$.

\textcircled{$\supseteq$}: proof by induction on the minimal number of steps
in the computation $p \rightarrowsymb{\epsilon/z} q$.

Induction base: if $p \rightarrowsymb{\epsilon/z} q$ is 
a single computation step (that is, a transition from the original relation) 
or less (that is, a trivial loop $p \rightarrowsymb{\epsilon/\epsilon} p$), 
then $R'$ contains the rule $\epsilon_{p,q} \to z$ by its construction.

Induction step: if $p \rightarrowsymb{\epsilon/z} q$ is computed in more than one step, 
it can be represented as a composition of  $p \rightarrowsymb{\epsilon/y_1} r$ and 
$r \rightarrowsymb{\epsilon/y_2} q$, where $z = y_1y_2$ and both sub-computations take fewer steps.
Then, by the induction hypothesis, $\str{x}{y_1} \in L_{G'}(\epsilon_{p,r})$ and 
$\str{xy_1}{y_2} \in L_{G'}(\epsilon_{r,q})$ for every $x \in \Omega^*$. 
Thus, applying the rule $\epsilon_{p,q} \to \epsilon_{p,r}\epsilon_{r,q}$
yields
$\str{x}{z} = \str{x}{y_1} \cdot \str{xy_1}{y_2} \in  L_{G'}(\epsilon_{p,q})$
for every $x \in \Omega^*$.
\end{proof}

\begin{lemma}\label{sublanguage_image_lemma}
Each nonterminal symbol $A_{p,q}$ in the constructed grammar,
with $A \in N$ and $p, q \in Q$,
defines the language
$L_{G'}(A_{p,q})$ of all strings $\str{x}{y}$,
with $x, y \in \Omega^*$,
such that there exists a string $\str{u}{v} \in L_G(A)$,
with $v \neq \epsilon$,
$q_0\rightarrowsymb{u/x} p$ and $p\rightarrowsymb{v/y} q$.
\end{lemma}

\begin{proof}
\textcircled{$\Rightarrow$}
Assume that $\str{x}{y} \in L_{G'}(A_{p,q})$;
it is claimed that there is a string $\str{u}{v} \in L_G(A)$,
where $v \neq \epsilon$,
$q_0\rightarrowsymb{u/x} p$ and $p\rightarrowsymb{v/y} q$.
The proof is by induction on the length of the derivation of $A_{p,q}(\str{x}{y})$ in $G'$
(assuming that all possible propositions of the form $\epsilon_{p',q'}(\str{x'}{y'})$ 
are already derived as per Lemma~\ref{epsilon_image_lemma}).

Induction base: if $A_{p,q}(\str{x}{y})$ is derived in one step, 
it must be derived by a rule of the form
$A_{p,q} \to \epsilon_{p,r}b\epsilon_{r',q} \& \before \epsilon_{q_0,p}$.
Then $\str{x}{y} = \str{x}{y_1by_2}$,
where $\str{\epsilon}{x} \in L_{G'}(\epsilon_{q_0,p})$,
$\str{x}{y_1} \in L_{G'}(\epsilon_{p,r})$
and $\str{x y_1}{y_2} \in L_{G'}(\epsilon_{r',q})$.
Then, by Lemma~\ref{epsilon_image_lemma},
$q_0\rightarrowsymb{\epsilon/x} p$, $p\rightarrowsymb{\epsilon/y_1} r$ and $r'\rightarrowsymb{\epsilon/y_2} q$.
Furthermore, the existence of such a rule implies existence of a rule $A\to a\&\before\epsilon$ in $G$, 
as well as a computation $r\rightarrowsymb{a/b} r'$.
Thus, taking $\str{u}{v} = \str{\epsilon}{a}$, we have $\str{\epsilon}{a} \in L_G(A)$, 
$q_0\rightarrowsymb{\epsilon/x} p$, and the composition of $p\rightarrowsymb{\epsilon/y_1} r$ with 
$r\rightarrowsymb{a/b} r'$ and $r'\rightarrowsymb{\epsilon/y_2} q$ produces $p\rightarrowsymb{a/y} q$.

Induction step: if $A_{p,q}(\str{x}{y})$ is derived in more than one step, 
it must be derived either by a rule of the form
$A_{p,q} \to \epsilon_{p,r}b\epsilon_{r',q} \& \before D_{q_0,p}$ with $D\in N$,
or by a rule of the form $A_{p,q} \to B^{(1)}_{p,r_1}C^{(1)}_{r_1,q}\&\ldots\& B^{(n)}_{p,r_n}C^{(n)}_{r_n,q}$.

Consider the first case: let $\str{x}{y}$ be a string in $L_{G'}(A_{p,q})$
that is derived by a rule $A_{p,q} \to \epsilon_{p,r}b\epsilon_{r',q} \& \before D_{q_0,p}$.
Then again $\str{x}{y} = \str{x}{y_1by_2}$,
where $\str{\epsilon}{x} \in L_{G'}(D_{q_0,p})$,
$\str{x}{y_1} \in L_{G'}(\epsilon_{p,r})$
and $\str{x y_1}{y_2} \in L_{G'}(\epsilon_{r',q})$.
By Lemma~\ref{epsilon_image_lemma},
$p\rightarrowsymb{\epsilon/y_1} r$ and $r'\rightarrowsymb{\epsilon/y_2} q$,
and additionally by the induction hypothesis there is such $\str{u_0}{v_0} \in L_G(D)$ that 
$q_0\rightarrowsymb{u_0/\epsilon} q_0$ and $q_0\rightarrowsymb{v_0/x}p$.
Then $u_0$ must be $\epsilon$, as otherwise the transducer would not be injective.
By the construction of the rule, there is such $a\in\Omega$ that the rule $A\to a\&\before D$ is in $R$,
and $r\rightarrowsymb{a/b}r'$. Let us prove that $\str{v_0}{a}$ satisfies the required condition on $\str{u}{v}$.
We already know that $q_0\rightarrowsymb{v_0/x}p$, and the composition of $p\rightarrowsymb{\epsilon/y_1} r$ with 
$r\rightarrowsymb{a/b} r'$ and $r'\rightarrowsymb{\epsilon/y_2} q$ produces $p\rightarrowsymb{a/y} q$.
It remains to apply the rule $A\to a\&\before D$ to $\str{v_0}{a}$, 
given $\str{\epsilon}{v_0} \in L_G(D)$.

The second case: let $\str{x}{y}$ be a string in $L_{G'}(A_{p,q})$
that is derived by a rule
$A_{p,q} \to B^{(1)}_{p,r_1}C^{(1)}_{r_1,q}\&\ldots\& B^{(n)}_{p,r_n}C^{(n)}_{r_n,q}$.
Then $\str{x}{y}$ is representable for each $i$ as a concatenation 
$\str{x}{y_i}\cdot\str{xy_i}{z_i}$ such that 
$\str{x}{y_i} \in L_{G'}(B^{(i)}_{p,r_i})$ and 
$\str{xy_i}{z_i} \in L_{G'}(C^{(i)}_{r_i,q})$.
By the induction hypothesis, it follows that there exist
$\str{u_i}{v_i} \in L_G(B^{(i)})$ and 
$\str{t_i}{w_i} \in L_G(C^{(i)})$ 
such that $q_0\rightarrowsymb{u_i/x}p$, 
$p\rightarrowsymb{v_i/y_i} r_i$, $q_0\rightarrowsymb{t_i/xy_i}r_i$,
$r_i\rightarrowsymb{w_i/z_i}q$.
By the injectivity of the transducer, all $u_i$ are equal,
$t_i = u_iv_i$, and all $v_iw_i$ are also equal.
Let $u = u_i$, $v = v_iw_i$. Then the original rule 
$A\to B^{(1)}C^{(1)}\&\ldots \&B^{(n)}C^{(n)}$ applies to 
$\str{u}{v} = \str{u_i}{v_i}\cdot\str{t_i}{w_i}$,
and $q_0\rightarrowsymb{u/x} p$, $p\rightarrowsymb{v/y}q$.

\textcircled{$\Leftarrow$}
Conversely, let $\str{u}{v} \in L_G(A)$,
where $v \neq \epsilon$,
and let $x, y\in \Omega^*$ be such that
$q_0\rightarrowsymb{u/x}p, p\rightarrowsymb{v/y}q$.
It is now claimed that $\str{x}{y} \in L_{G'}(A_{p,q})$.
This time the proof is given by induction
on the length of the derivation of $A(\str{u}{v})$ in $G$.

Induction base: If $A(\str{u}{v})$ is derived in one step, 
it must be derived by a rule of the form $A\to a\&\before\epsilon$.
Then $u = \epsilon, v = a$, and also $q_0\rightarrowsymb{\epsilon/x}p, p\rightarrowsymb{a/y}q$.
The latter can be decomposed as 
$p\rightarrowsymb{\epsilon/y_1}r, r\rightarrowsymb{a/b}r', r'\rightarrowsymb{\epsilon/y_2}q$ 
for some $y_1, y_2 \in \Omega^*, b\in\Omega\cup\{\epsilon\}$. 
The rule $A_{p,q} \to \epsilon_{p,r}b\epsilon_{r',q} \& \before \epsilon_{q_0,p}$ can then be applied to derive 
$A_{p,q}(\str{x}{y})$.

Induction step: again, we have to consider two cases.
First case: let $\str{u}{a}$ be a string in $L_G(A)$
that is derived by a rule $A \to a\&\before D$, and let
$q_0\rightarrowsymb{u/x}p, p\rightarrowsymb{a/y}q$.
Then $y = y_1by_2$, where
$\epsilon_{p,r}(\str{x}{y_1}), r\rightarrowsymb{a/b}r',
\epsilon_{r',q}(\str{xy_1b}{y_2})$.
It remains to prove that $\str{\epsilon}{x} \in L_{G'}(D_{q_0,p})$,
which by the induction hypothesis follows from existence of some
$\str{u'}{v'}\in L_G(D)$ such that $q_0\rightarrowsymb{u'/\epsilon}q_0$ 
and $q_0\rightarrowsymb{v'/x}p$.
Such a witness can be easily found as $u'\langle v'\rangle = \epsilon\langle u\rangle$.

Second case: let $\str{u}{v}$ be a string in $L_G(A)$
that is derived by a rule $A \to B^{(1)}C^{(1)}\&\ldots \&B^{(n)}C^{(n)}$, and let
$q_0\rightarrowsymb{u/x}p, p\rightarrowsymb{v/y}q$.
Then for each $i$ there is a decomposition of $v = v_iw_i$
with $\str{u}{v_i} \in L_G(B^{(i)})$ and $\str{uv_i}{w_i} \in L_G(C^{(i)})$.
As the computation $p\rightarrowsymb{v/y}q$ reads at most one symbol at each step,
there is some transitional state $r_i$ that allows for its decomposition as 
$p\rightarrowsymb{v_i/y_i}r_i, r_i\rightarrowsymb{w_i/z_i}q$ with $y_iz_i = y$.
Then by the induction hypothesis
$\str{x}{y_i} \in L_{G'}(B^{(i)}_{p,r_i})$ and $\str{xy_i}{z_i}\in L_{G'}(C^{(i)}_{r_i,q})$,
which allow for derivation of $A_{p,q}(\str{x}{y})$ by the rule
$B^{(1)}_{p,r_1}C^{(1)}_{r_1,q}\&\ldots\& B^{(n)}_{p,r_n}C^{(n)}_{r_n,q}$.

\end{proof}

Now we can use Lemma~\ref{sublanguage_image_lemma} to prove the theorem.

\begin{proof}[Proof of Theorem~\ref{injective_image_theorem}]
Let $u \in L(G')$.
This is equivalent to either $\str{\epsilon}{u} \in L_{G'}(S_{q_0, q_F})$,
or $\str{\epsilon}{u} \in L_{G'}(\epsilon_{q_0, q_F})$
if $G$ has the rule $S \to \epsilon$.
In turn, the first option is equivalent to the existence of $\str{x}{y} \in L_G(S), y\ne\epsilon$
such that $q_0\rightarrowsymb{x/\epsilon}q_0$ (which, again, means that $x = \epsilon$) 
and $q_0\rightarrowsymb{y/u}q_F$,
while the second option is equivalent to the existence of $\str{\epsilon}{\epsilon} \in L_G(S)$
and $q_0\rightarrowsymb{\epsilon/u}q_F$.
Notice that the nonemptiness clause in the first option is covered by the second option, 
so the original statement is simply equivalent to the existence of $y$
such that $\str{\epsilon}{y} \in L_G(S)$ (which is the definition of $y \in L(G)$)
and $q_0\rightarrowsymb{y/u}q_F$ (which is the definition of $u \in \mathcal{T}(y)$).
\end{proof}

It should be noted that the definition of an NFT is symmetric with respect to the input and the output.
Let $\mathcal{T}'$ be the inverse NFT
derived from $\mathcal{T}$ by swapping $\Sigma$ and $\Omega$,
and accordingly replacing each transition $(p, u, x, q)$ in $\mathcal{T}$
with a transition $(p, x, u, q)$.
Then $x \in \mathcal{T}(u) \Leftrightarrow u \in \mathcal{T}'(x)$.
Furthermore, for an injective NFT,
its inverse is a NFT that maps each string to at most one string,
and thus implements a function.
This closure therefore implies the closure under inverse homomorphisms,
as they are a special case of such functions.

\begin{corollary}
A language is defined by a grammar with left context operators
if and only if
it is representable as an inverse homomorphic image of $L_0$.
\end{corollary}

\section{Conclusion}

A subject suggested for future research
is investigating whether
\emph{grammars with two-sided context operators}~\cite{ContextsTwoSided}
have a hardest language.
All that is known about these grammars
is a basic normal form theorem~\cite{ContextsTwoSided}
and a cubic-time parsing algorithm~\cite{Rabkin}.
The methods of the present paper might still be applicable
to constructing a hardest language;
however, this would likely require developing more sophisticated normal forms first.

Another related problem is the existence of a hardest language
for \emph{linear grammars with left context operators}~\cite{ContextsLinear}.
Whether the methods recently used to prove that there is no hardest language
for the related family of \emph{linear conjunctive grammars}~\cite{MrykhinOkhotin_cellular}
would apply in this case, remains to be seen.


\begin{thebibliography}{99}

\bibitem{Autebert_counter_invh} J.-M. Autebert,
	\href{http://dx.doi.org/10.1007/BF01768474}
	{``Non-principalit\'e du cylindre des langages \`a compteur''},
	\emph{Mathematical Systems Theory},
	11:1 (1977), 157--167.

\bibitem{GrammarsWithContexts} M. Barash, A. Okhotin,
	\href{http://dx.doi.org/10.1016/j.ic.2014.03.003}
	{``An extension of context-free grammars with one-sided context specifications''},
	\emph{Information and Computation},
	237 (2014), 268--293.

\bibitem{ContextsTwoSided} M. Barash, A. Okhotin,
	\href{http://dx.doi.org/10.1016/j.tcs.2015.05.004}
	{``Two-sided context specifications in formal grammars''},
	\emph{Theoretical Computer Science},
	591 (2015), 134--153.

\bibitem{ContextsLinear} M. Barash, A. Okhotin,
	\href{http://dx.doi.org/10.1051/ita/2015004}
	{``Linear grammars with one-sided contexts and their automaton representation''},
	\emph{RAIRO Informatique Th\'eorique et Applications},
	49:2 (2015), 153--178.

\bibitem{ContextsLR} M. Barash, A. Okhotin,
	{``Generalized LR parsing for grammars with contexts''},
	\emph{Theory of Computing Systems},
	61:2 (2017), 581--605.

\bibitem{ContextsLinearSpace} M. Barash, A. Okhotin,
	\href{https://doi.org/10.1016/j.tcs.2017.11.006}
	{``Linear-space recognition for grammars with contexts''},
	\emph{Theoretical Computer Science},
	719 (2018), 73--85.

\bibitem{BoassonNivat} L. Boasson, M. Nivat, 
	\href{https://doi.org/10.1007/BF01768473}
	{``Le cylindre des langages linéaires``}, 
	\emph{Mathematical Systems Theory}, 
	11 (1977), 147--155.

\bibitem{CulikMaurer} K. \v{C}ul\'{\i}k II, H. A. Maurer,
	\href{http://eudml.org/doc/92102}
	{``On simple representations of language families''},
	\emph{RAIRO Informatique Th\'eorique et Applications},
	13:3 (1979), 241--250.

\bibitem{ElgotMezei} C. C. Elgot, J. E. Mezei,
	\href{http://dx.doi.org/10.1147/rd.91.0047}
	{``On relations defined by generalized finite automata''},
	\emph{IBM Journal of Research and Development},
	9:1 (1965), 47--68.

\bibitem{GinsburgUllian_preservation} S. Ginsburg, J. Ullian,
	\href{http://dx.doi.org/10.1145/321341.321345}
	{``Preservation of unambiguity and inherent ambiguity in context-free languages''},
	\emph{Journal of the ACM},
	13:3 (1966), 364--368.

\bibitem{Greibach} S. A. Greibach,
	\href{https://doi.org/10.1137/0202025}
	{``The Hardest Context-Free Language``},
	\emph{SIAM Journal on Computing},
	2(4), 304--310.
	
\bibitem{Greibach_jump} S. A. Greibach,
	\href{http://dx.doi.org/10.1137/0203009}
	{``Jump PDA's and hierarchies of deterministic context-free languages''},
	\emph{SIAM Journal on Computing},
	3:2 (1974), 111--127.

\bibitem{BooleanInvGSM} T. Lehtinen, A. Okhotin,
        \href{http://dx.doi.org/10.1142/S0129054110007568}
        {``Boolean grammars and GSM mappings''},
        \emph{International Journal of Foundations of Computer Science},
        21:5 (2010), 799--815.

\bibitem{MrykhinOkhotin_cellular} M. Mrykhin, A. Okhotin,
	\href{https://doi.org/10.1007/978-3-030-68195-1_10}
	{``On hardest languages for one-dimensional cellular automata''},
	\emph{Language and Automata Theory and Applications}
	(LATA 2021, Milan, Italy, 1--5 March 2021),
	LNCS 12638, 118--130.

\bibitem{MrykhinOkhotin_LL} M. Mrykhin, A. Okhotin,
	\href{https://doi.org/10.1007/978-3-030-81508-0_25}
	{``The hardest LL($k$) language''},
	\emph{Developments in Language Theory}
	(DLT 2021, Porto, Portugal, 16--20 August 2021),
	LNCS 12811, 304--315.

\bibitem{Conjunctive} A. Okhotin,
	``Conjunctive grammars'',
	\emph{Journal of Automata, Languages and Combinatorics},
	6:4 (2001), 519--535.

\bibitem{BooleanMatrix} A. Okhotin,
	\href{http://dx.doi.org/10.1016/j.tcs.2013.09.011}
	{``Parsing by matrix multiplication generalized to Boolean grammars''},
	\emph{Theoretical Computer Science},
	516 (2014), 101--120.

\bibitem{ContextsImproved} A. Okhotin,
	\href{http://dx.doi.org/10.1016/j.tcs.2015.03.041}
	{``Improved normal form for grammars with one-sided contexts''},
	\emph{Theoretical Computer Science},
	588 (2015), 52--72.

\bibitem{ConjunctiveTokyo} A. Okhotin,
	\href{https://doi.org/10.1007/978-3-319-98653-1_4}
	{``A tale of conjunctive grammars''},
	\emph{Developments in Language Theory}
	(DLT 2018, Tokyo, Japan, 10--14 September 2018),
	LNCS 11088, 36--59.

\bibitem{ConjunctiveHardest} A. Okhotin,
	\href{https://doi.org/10.1016/j.ic.2018.11.001}
	{``Hardest languages for conjunctive and Boolean grammars''},
	\emph{Information and Computation},
	266 (2019), 1--18.

\bibitem{ConjunctiveNoUnion} A. Okhotin, C. Reitwie{\ss}ner,
	\href{http://dx.doi.org/10.1016/j.tcs.2010.03.015}
	{``Conjunctive grammars with restricted disjunction''},
	\emph{Theoretical Computer Science},
	411:26--28 (2010), 2559--2571.

\bibitem{Rabkin} M. Rabkin,
	\href{http://dx.doi.org/10.1007/978-3-319-06686-8_24}
	{``Recognizing two-sided contexts in cubic time''},
	\emph{Computer Science---Theory and Applications}
	(CSR 2014, Moscow, Russia, 6--12 June 2014),
	LNCS 8476, 314--324.

\end{thebibliography}
\end{document}